\documentclass[twocolumn,secnumarabic,amssymb,amsmath, nobibnotes, aps,showpacs,superscriptaddress,prd,floatfix]{revtex4-1}
\usepackage{color}
\usepackage{amssymb,amsmath,graphicx}
\usepackage{lineno}

\begin{document}


\title{ Cross Section for Rydberg Antihydrogen Production via Charge Exchange Between Rydberg Positronium and  Antiprotons in Magnetic Field}
\author{D. ~Krasnicky}
\affiliation{Istituto Nazionale di Fisica Nucleare (INFN) Genoa and Department of Physics, University of Genoa
 via Dodecaneso 33, 16146 Genoa, Italy}
\author{R. ~Caravita}
\affiliation{Istituto Nazionale di Fisica Nucleare (INFN) Genoa and Department of Physics, University of Genoa
 via Dodecaneso 33, 16146 Genoa, Italy}
\author{C. Canali }
\affiliation{Istituto Italiano di Tecnologia Via Morego 30, 16163 Genoa, Italy}
\author{G.~Testera}
\affiliation{Istituto Nazionale di Fisica Nucleare (INFN)  Genoa, via Dodecaneso 33, 16146 Genoa, Italy}

\begin{abstract}

{The antihydrogen formation by charge exchange between cold antiprotons and Rydberg positronium $P_s^*$ is studied by using the Classical Trajectory Monte Carlo (CTMC) method.
In absence of external magnetic field the cross section scaled by the fourth power of the $P_s^*$ principal quantum number  $n_{P_s}$ shows an universal behaviour as a function of the ratio $k_v$ between the velocity of the $P_s$ centre of mass and that of the 
positron in the classical circular orbit. At low velocity, below about  $k_v \simeq 0.2- 0.3$,  we show for the first time for Rydberg positronium, that the cross section  increases as  $1/k_v^2$ or, in equivalent way  as $1/E_{Ps}^{cm}$ with $E_{Ps}^{cm}$ being the $P_s^*$ centre of mass energy. 
In this regime the distribution of the principal quantum number of the antihydrogen state is narrow at peaked around $\sqrt{2²} n_{P_s}$ while at higher $k_v$ values
a broad distribution of antihydrogen  states is produced. 
The study of the   collision process in presence of moderate magnetic field (0.5-2 T) shows that there is an experimentally interesting region of $k_v$ with the cross section  slightly higher than that in absence of field. 
However the presence of a magnetic field changes significantly the cross section behaviour as a function of $k_v$, especially at low velocities, where 
reductions of the cross sections and deviations from the $1/k_v^2$  ($1/E_{Ps}^{cm}$) are observed. 
Our calculation shows for the first time a dependance of the cross section upon the angle between the magnetic field and the flight direction of the incoming $P_s^*$. }

\end{abstract}



\maketitle




\section{Introduction}
\label{sec:intro}
Antihydrogen atoms are a powerful physical system for accurate tests of some of the  fundamental principles  of physics.
The precision measurement of the transition energies of the antihydrogen levels (specially the 1S-2S line or the hyperfine splitting of the fundamental state)
and the comparison with the corresponding ones of hydrogen could result in the most precise  test of the CPT symmetry for baryons ever performed \cite{Widmann}. In addition the direct measurement of the 
Earth's gravitational acceleration g on antihydrogen would allow to probe the validity of the weak equivalence principle (WEP)  for a system made only by antimatter
\cite{AEgIS}.
These two principles are related to the foundations of quantum field theory (CPT) and of General Relativity (WEP) and presently 
none has ever observed a CPT violation, nor a process in which WEP is not satisfied \cite{PDG}.
However the search for any possible tiny violation is of interest as its discovery would be a signal of new  physics \cite{NonStandard}. 

The formidable accuracies reached in the field of hydrogen spectroscopy \cite{1S2S} and gravitational measurements with cold atoms \cite{ColdAtoms} represent the long term goal of the antihydrogen experiments. Presently there is still 
an experimental gap to be bridged between the cold atom and cold antiatom physics which is dominated primary by the different values of the temperature of the available samples but also by the difference in the number of available particles.
While 
ordinary atoms are at one's disposal in large quantities and can be cooled to $\mu K$ or $nK$  temperature, only small numbers of antihydrogen atoms are presently produced  with temperature in the range of K \cite{HighRate}.
 High sensitivity spectroscopy and precision gravitational measurements on antihydrogen  both demand  to prepare  antiatoms with sub-Kelvin temperature, possibly  in the range of mK or below.
 
After the production of antihydrogen with a temperature of about some tens of Kelvin by the ATHENA \cite{Athena} and ATRAP \cite{Atrap} experiments in 2002,
the present challenge of the on going experimental activity is towards the production of antiatoms as cold as possible and in large quantities.
 The  efforts  are focused on trapping antihydrogen in a magnetic trap as in the ALPHA \cite{Alpha} or ATRAP \cite{AtrapTrapping}  experiments or producing
a cold beam as in the AEgIS \cite{AEgIS} or ASACUSA \cite{ASACUSA} ones or, finally, on getting cold antihydrogen through the intermediate formation of charged antimatter ions  as  in the GBAR project \cite{GBAR}.

Antihydrogen atoms are  produced by three-body recombination of antiprotons and positrons trapped and cooled in electromagnetic traps \cite{3body} or by charge exchange between Rydberg positronium ($P_s^*$) and antiprotons. 
The last reaction 
\begin{equation}
P_s^* +\bar p \rightarrow \bar H^* + e^-
\label{eq:1}
\end{equation}
pioneered by the ATRAP collaboration \cite{AtrapCC}, is the main antihydrogen formation mechanism in the AEgIS experiment. 

In this paper we present detailed results about the calculation of the charge exchange cross section 
obtained with  a Classical Trajectory Montecarlo method (CTMC).
We assumed that antiprotons are at rest and we studied the collision process as a function of the positronium centre of mass velocity 
for various 
 principal quantum number of the positronium $n_{P_s}$.
We first performed the calculation in absence of magnetic field and then we included the effect of moderate B values (around 1-2 Tesla) as used in the AEgIS experiment.
Previous works concerning charge exchange of antiprotons  with Rydberg positronium are limited to  collision velocities and magnetic fields higher than that considered here \cite{Lu}
and to a different dynamic regime in which  the initial positronium state is a long-lived delocalized outer well state \cite{bib:PsDelocalized}.
  Other works \cite{Charlton1}, \cite{Charlton2} extend the calculation to to low collision velocity but only consider positronium in low excited states ($n_{P_s}$=3 at maximum). Finally other studies \cite{Homan}  
  \cite{TwoStage}
are focused on modelling the dynamics of the antihydrogen formation by a double process of charge exchange (as in the ATRAP experiment \cite{AtrapCC}), the first one producing the Rydberg positronium and the second one producing the antihydrogen.

Our calculation   show that, in absence of magnetic fields, the cross section scaled by $n_{P_s}^4$ has an universal shape as a function of 
the ratio  $k_v$ between the velocity of the centre of mass of $P_s^*$ and that of the positron in the positronium classical circular orbit.
This universal shape is valid for all the values of  $n_{P_s}$ that we have investigated (ranging from 3 to 50).
Below about $k_v \simeq 0.3 $  the scaled cross section  increases 
as $1/ k_v^2$  or, in equivalent way, 
as  $1/ E_{P_s}^{cm}$ being  $E_{P_s}^{cm}$ the 
  positronium centre of mass  kinetic energy. This result extends the one  reported in  \cite{Charlton1}, \cite{Charlton2} limited to  $n_{P_s} \leq 3$.
This low velocity behaviour of the charge exchange process and the high values of the cross section have interesting consequences for the antihydrogen experiments and suggest that the production of a large number of cold antihydrogen needs  very  cold Rydberg  positronium.

The interest   of the reaction of antiprotons  with cold positronium emerges also from the analysis of the distribution of the principal  quantum number of the formed antihydrogen:
 low velocity collisions (in the $1/ E_{P_s}^{cm}$  regime) produce antihydrogen with a narrow distribution of principal quantum numbers which is advantageous for performing on them  further atomic manipulations 
 \cite{TesteraAIP}.
 Higher velocity collisions produce antihydrogen populating a distribution with a large spread  of principal quantum numbers.

 We included in the calculation the 
  presence of a magnetic field  as needed to trap the antiprotons: we found that  at very low velocities the cross section no longer increases as $1/ k_v^2$  ($1/E_{P_s}^{cm}$)
   and its universal behaviour  is broken. 
This reduction of the cross section depends on the flight direction of the incoming positronium with respect to the magnetic field.
Interestingly there is a significant  range of collision velocity where the cross section increases in presence of magnetic field. 

We first recall some basic principle of the CTMC method and we show the results obtained in absence of magnetic field. We then explain how the CTMC method is extended to include the effect of the magnetic field and we show the relevant results.

\section{Classical Trajectory Monte Carlo Method }
\label{CTMCDescription}
In absence of magnetic field the interaction  between $P_s^*$ in the initial quantum state defined by $n_{P_s},l_{P_s},m_{P_s}$ with an antiproton  may result in the antihydrogen formation (see equation \ref{eq:1}) but also in  $P_s^*$ elastic or inelastic scattering or ionisation as relation  \ref{eq:collisione} shows.

\begin{equation}
\left \{ \begin{array}{lcccccl}
P_s^* (n_{P_s},l_{P_s},m_{P_s}) +\bar p \rightarrow P_s^* (n_{P_s},l_{P_s},m_{P_s}) + \bar p \\
\\
P_s^* (n_{P_s},l_{P_s},m_{P_s}) +\bar p \rightarrow P_s^* (n_{P_s}^{\prime},l_{P_s}^{\prime},m_{P_s}^{\prime}) + \bar p \\
\\
P_s^* (n_{P_s},l_{P_s},m_{P_s}) +\bar p \rightarrow e^+ + e^- + \bar p
  \end{array}  \right.
\label{eq:collisione}
\end{equation}

The CTMC method was introduced in \cite{AbrinesPercival}
 to calculate capture and ionisation cross sections for
proton-hydrogen collisions \cite{pHCollisions} and it has been extensively used also to model three-body processes \cite{3bodyOlson} and multielectrons targets   \cite{Handbook}.
It is particularly well suited to  model processes involving Rydberg atoms and automatically allows to account for all the mentioned collision channels.

The CTMC procedure is based on solving the classical equation of motion for a a three-body, three-dimensional system made off the positron and electron initially bound in the positronium  atom and the target antiproton.  
We solve the classical equation of motion with  the hamiltonian $H_{tot}$  
neglecting the spin. Using atomic units, $H_{tot}$ in absence of external fields  is given by 
\begin{equation}
H_{tot} = \frac{\pi_{\bar p}^2}{2m_{\bar p}} + \frac{\pi_{e^+}^2}{2} + \frac{\pi_{e^-}^2}{2} - \frac{1}{r_{e^+e^-}} - \frac{1}{r_{\bar p e^+}} + \frac{1}{r_{\bar p e^-}}
 \label{eq:Htot}
\end{equation}
where $\vec r_{\bar p}$ and $\vec \pi_{\bar p}= m_{\bar p} \vec v_{\bar p}$ are the position and the mechanical momentum of the antiproton in the laboratory reference frame,
$\vec r_{e^+}$, $\vec \pi_{e^+}$,  $\vec r_{e^-}$ and $\vec \pi_{e^-}$  are the corresponding quantities for the positron and the electron
and $r_{e^+e^-}$, $r_{{\bar p}e^+}$, $r_{{\bar p}e^-}$ are the distances between each couple of particles.

The initial conditions are randomly sampled. For each sorted initial state the classical trajectories are calculated starting from a large separation between antiproton and positronium to a distance of closest approach and out again to a large separation between the antiproton and the electron.
The Coulomb force among the three-body is included in all the steps of the simulation.
If at the end of each simulation the positron is found to be bound to the antiproton then the reaction is classified as antihydrogen formation.
In detail the steps of the CTMC method are: 
\begin{itemize}
\item sampling of the initial conditions;
\item integration of the equations of motion;
\item identification of the final conditions;
\item calculation of the cross section.
\end{itemize}

\subsection{Sampling of the initial conditions}
\label{sec:initial}
\begin{figure}
\begin{center}
\hspace{-5 mm}
\centering{\includegraphics[width = 0.5 \textwidth]{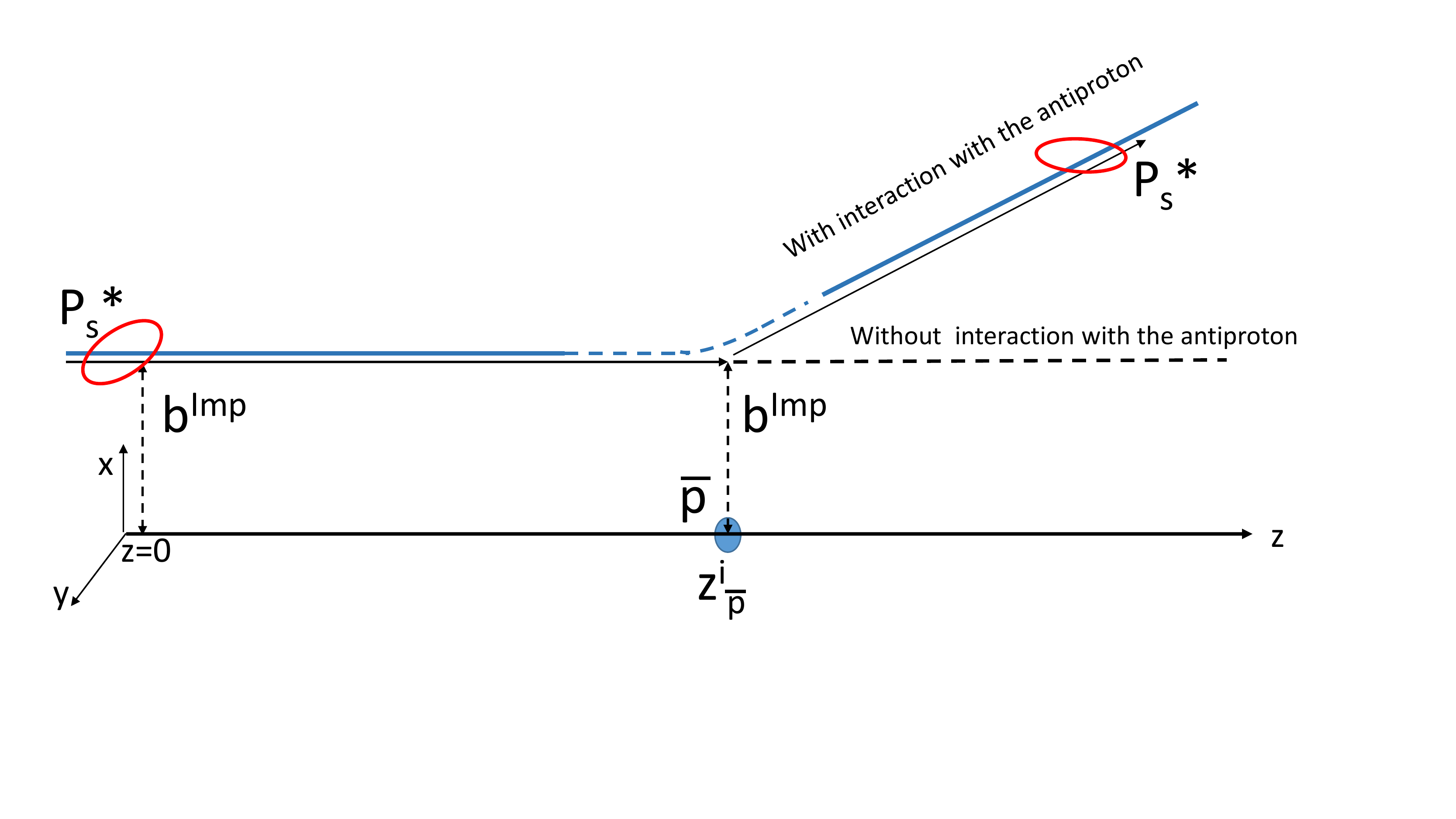}}
\caption{Definition of the geometry of the collision without magnetic field with a pictorial view of the trajectories of the centre of mass of $P_s^*$ with and without interaction with the antiproton. The trajectory of  $P_s^*$ in case of absence of interaction with the antiproton is a straight line along $z$. 
$b^{Imp}$ is the impact parameter.
 }
\label{fig:ImpactNoB}
\end{center}
\end{figure}     

Figure \ref{fig:ImpactNoB} shows the geometry of the collision. 
The antiproton is initially at rest in the position $z=z_{\bar p}^i$.
The initial  position of the centre of mass of the $P_s^*$ atom is in the z=0 plane; the impact parameter $b^{Imp}$ is chosen generating its radial position within  a circle of radius $b_{max}$. 
The value of $b_{max}$ may depend on the process in which we are  interested (ionisation, excitation, charge exchange): it is chosen  as the minimum value $b_{max}$ such  that
adding trajectories with 
  $b^{Imp}>b_{max}$ the resulting variation of the cross section is negligible within the statistical uncertainty of the calculation.

The initial conditions in the phase space describing the positronium must be selected from  a statistical distribution of the classical variables that matches  the corresponding quantum mechanical distribution. 
As widely discussed in \cite{AbrinesPercival} \cite{McKellar1984}, we adopt the choice of picking up  initial conditions from a microcanonical ensemble. This allows matching
the quantum mechanical energy and momentum distributions.
The generation of the initial conditions for $P_s^*$ begins by considering the Hamiltonian of the $e^+$, $e^-$ system and separating the centre of mass motion from the relative motion of the $e^+$ and $e^-$.
The relative motion is that of a particle with reduced mass $\mu=1/2$  in the Coulomb potential and the classical orbits for the bound system are Kepler elliptical orbits. Energy and angular momentum are conserved.

$P_s^*$ in the quantum state with principal quantum number $n_{P_s}$ is then described by generating ellipses corresponding to the energy $E_{n_{P_s}}= \dfrac {1}{4 n_{P_s}^2}$. Specifying the energy only  defines the semi-axis major $a_{n_{P_s}}$ ($a_{n_{P_s}} = 2 n_{P_s}^2$)
being the semi-axis minor $b_{{nl}_{P_s}}$ related to the classical angular momentum $L_c$. In the micro-canonical ensemble \cite{AbrinesPercival} the classical squared angular momentum 
$L_c^2= (\vec r \wedge \vec \pi)^2$ is uniformly distributed between 0 and its maximum allowed value.
For a given $n_{P_s}$ value we then generated  $L_c^2$  with uniform distribution and then the corresponding $b_{{nl}_{P_s}}$. The quantal weights are reproduced for all the $l_{P_s}$ values by defining 
$l_{P_s}$ as 

\begin{equation}
l_{P_s}\ll L_c/\hbar \le l_{P_s}+1   
\label{eq:ldef}
\end{equation}
with $l_{P_s}=0,1,...n_{P_s}-1$.

$b_{{nl}_{P_s}}$ is then given by $b_{{nl}_{P_s}} = 2 n_{P_s} \sqrt{{l_{P_s}}(l_{P_s}+1)}$.

The starting coordinates and velocities   in the ellipse have been  generated  by solving the motion equation of the selected elliptical trajectory  for  one period and then picking up  a time value 
with random uniform distribution between 0 and the ellipse period (and the corresponding coordinates  and velocities).

The orientation of the ellipse plane  is linked to the projection of the angular momentum along the $z$ axis and thus to the $m_{P_s}$ quantum number.
All the $m_{P_s}$ values are generated by introducing a rotation with three Euler angles.

The velocity of the centre of mass of the positronium $\vec v_{P_s}^{cm}$ defines the collision velocity as we assume   that the antiproton is at rest. 
Of course identical results would be   obtained by considering the motion of the antiproton and defining $\vec v_{P_s}^{cm}$ as the relative velocity (in the laboratory frame) between positronium and antiproton.
It is generally known that  the processes corresponding  to relations  \ref{eq:1}, \ref{eq:collisione} involving Rydberg atoms and ions \cite{McKellar1984} have huge cross sections when the impact speed is close to 
the mean speed $v_n$ of the Rydberg electron. 
 We thus define the parameter $k_v$ as the ratio between $v_{P_s}^{cm}$  and the velocity of the 
 positron $\dfrac{1}{2n_{P_s}}$ in the  $P_s^*$ centre of mass in the circular orbit
 \begin{equation}
 k_v = \dfrac{v_{P_s}^{cm}}{2n_{P_s}}
 \label{eq:kv}
 \end{equation}
 We assumed that  $\vec v_{P_s}^{cm}$ is  along the $z$ direction.
\subsection{Integration of the equations of motion}
We used a six order Runge Kutta method with a variable time interval.
We calculated at each step $H_{tot}$ and we used the difference $H_{diff}$ between the actual value of $H_{tot}$ and its initial value as a check of the accuracy  of the calculation.
Typically $\mid H_{diff}/H_{tot} \mid \simeq  10^{-7}$. Trajectories that do not conserve the energy were discarded. They are less than 0.05 $\%$ of the total.

We selected  $z_{\bar p}^i$ (see figure \ref{fig:ImpactNoB})   and the distance   between the antiproton and the electron where the simulation should be stopped
three times larger  than the maximum impact parameter. We have checked the stability of the results
with respect to these choices.

\subsection{Identification of the final conditions}
The classification  of the final state is performed analysing the Hamiltonian $H_{e^+ e-}$, $H_{\bar p e^+}$ of the relative motion between couples of particles.
\begin{equation}
H_{e^+ e-} = \pi_{e^+ e^-} ^2- \frac{1}{r_{e^+e^-}}
 \label{eq:HPs}
\end{equation}
 where $\pi_{e^+ e^-} = \pi_{e^-} -\pi_{e^+}$ is the mechanical momentum of the relative motion of the couple $e^+ e-$
 \begin{equation}
H_ {\bar p e^+}  = \frac {\pi_{\bar p e^+}^2}{2 m_{\bar p}}- \frac{1}{r_{\bar p e^+}}
 \label{eq:HantiH}
\end{equation}
and $\pi_{\bar p e^+} = \pi_{e^+} -\pi_{\bar p}$

If at the end of the collision $H_{e^+ e-} <0$ and $H_{\bar p e^+} >0$  then  the electron and positron  are still bound into the positronium. 
The principal quantum number $n_{P_s}^\prime$ is defined according to the value of $H_{e^+ e-}$ and 
the comparison between $n_{P_s}$ and $n_{P_s}^\prime$ establishes if the collision is elastic or inelastic.
If $H_{e^+ e-} >0$ and $H_{\bar p e^+} >0$ then positronium is ionised. Finally if
$H_{e^+ e-} >0$ and $H_{\bar p e^+} <0$  the positron is bound to the antiproton,  the collision is classified as antihydrogen formation and its quantum numbers are evaluated.
 
\subsection{Calculation of the cross section}

The cross section in SI units for charge exchange $\sigma$ and its standard (r.m.s.) error  $\Delta \sigma$ are obtained using \cite{AbrinesPercival}
\begin{equation}
\sigma= \pi a_0^2  b_{max}^2 \dfrac{N_{\bar H }}{N_{tot}}
\label{eq:Xsec}
\end{equation}

\begin{equation}
\Delta \sigma= \sigma \sqrt{\dfrac {N_{tot}-N_{\bar H }}{N_{tot} N_{\bar H }}}
\label{eq:XsecErr}
\end{equation}
where $b_{max}$ is the maximum value of the impact parameter in atomic units;
 $N_{\bar H }$ is the number of trajectories resulting in anti hydrogen formation and $N_{tot}$ is the total number of generated trajectories.

The statistical uncertainty  of each point in the cross section plots shown along this paper is often  hidden within the size of the plot markers. Typically we run a number of trajectories sufficient to calculate the cross section with a statistical error of $2-3 \%$ in absence of magnetic field. In presence of magnetic field, being the computation time longer, in some case  the statistical accuracy
 is  smaller as it appears  in the plots.
The number of trajectories to be generated depends on the parameters of the collision and it is typically of the order of several  tens of thoushand.

\section{Charge exchange cross section in absence of magnetic field}

\begin{figure*}
\begin{center}
\hspace{-5 mm}
\centering{\includegraphics[width = 1.0 \textwidth]{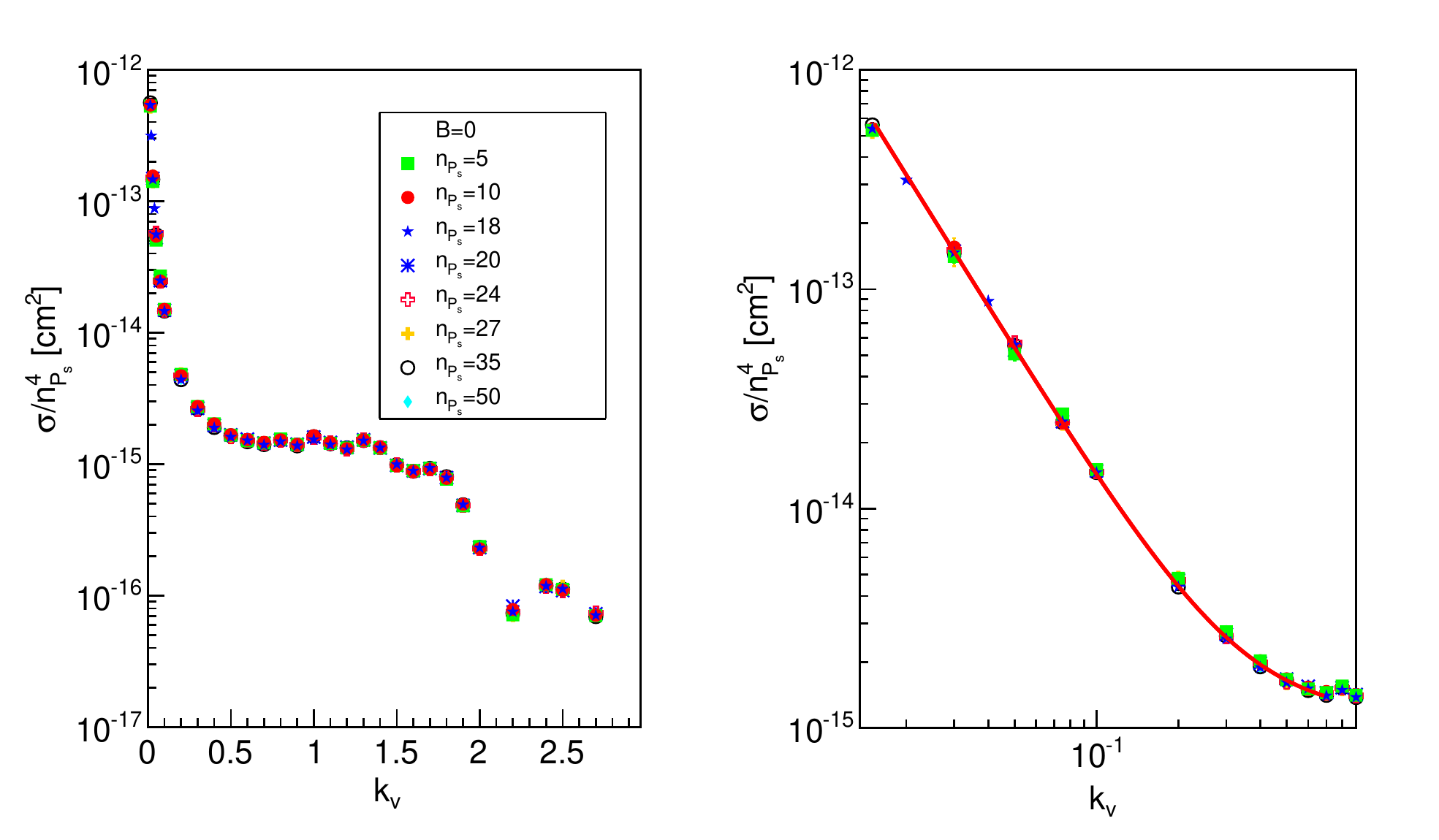}}
\caption{Charge exchange cross section divided by $n_{Ps}^4$ ($\sigma/n_{Ps}^4$) as a function of $k_v$ with B=0.
The results obtained  for the various principal quantum number shown in the legend collapse into a universal curve and they  cannot be distinguished in the plot. For each $n_{P_s}$ the $l_{P_s}$ and $m_{P_s}$  values are sampled from a canonical ensemble as described in section \ref{CTMCDescription}. The right plot is a zoom of the region with low $k_v$ values with the fit  
$ \sigma/n_{P_s}^4 [cm^2]= \dfrac{s_1}{k_v^2}+ s_2 $ superimposed (red line). $s_1=1.32 \cdot 10^{-16} cm^2$, $s_2= 1.12 \cdot 10^{-15} cm^2$.
 }
\label{fig:AllB0}
\end{center}
\end{figure*}

\begin{figure}
\begin{center}
\hspace{-5 mm}
\centering{\includegraphics[width = 0.5\textwidth]{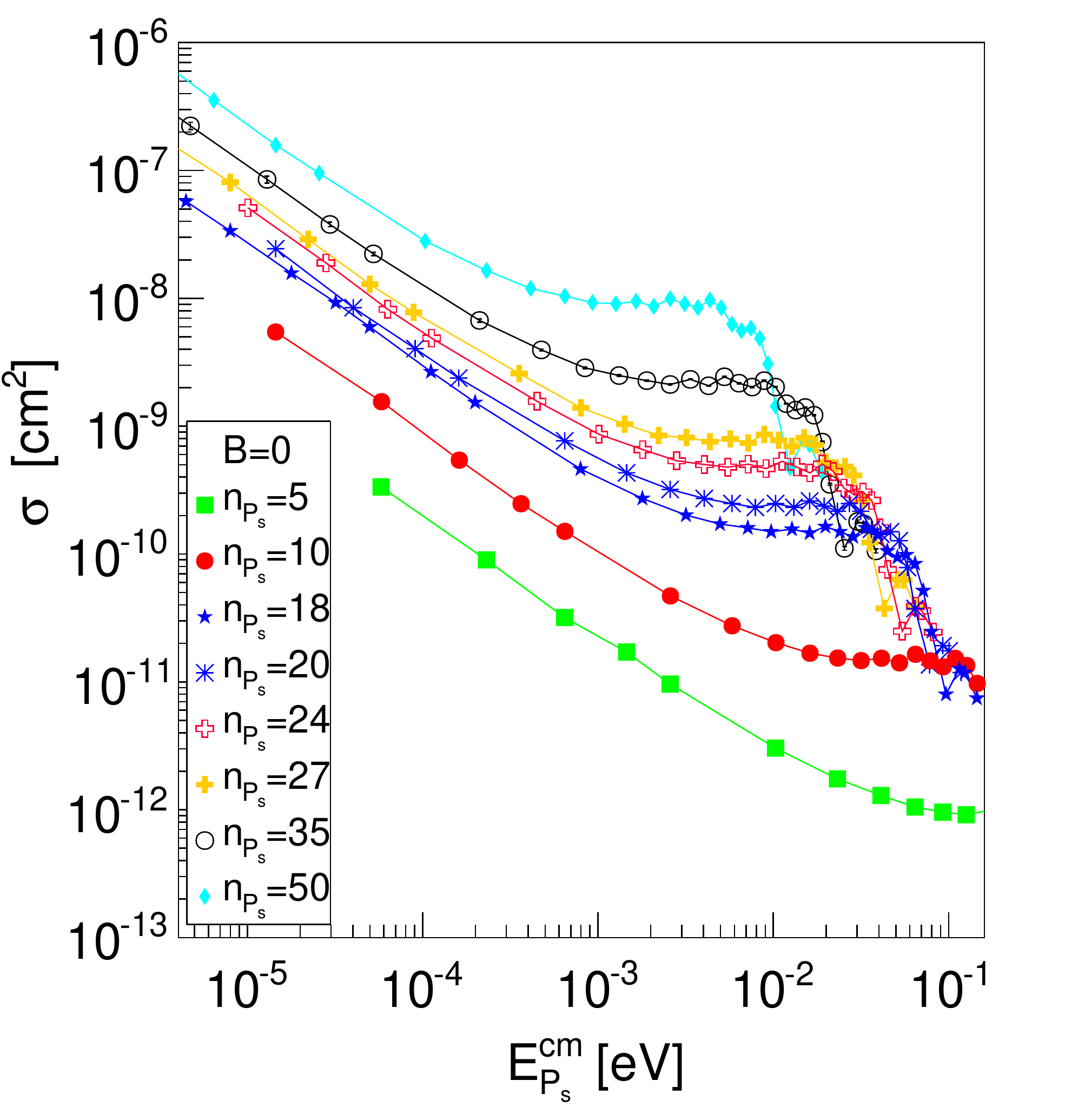}}
\caption{Charge exchange cross section $\sigma$ as a function of the Ps centre of mass energy. The plot shows the same points of figure \ref{fig:AllB0}.
The lines simply connect the points to help the  graphical interpretation.}
\label{fig:AllB0Ene}
\end{center} 
\end{figure}    

We first considered positronium in a initial state with $n_{P_s}$  defined and with all possible values of $l_{Ps}$ and $m_{Ps}$ (distributed as described in section \ref{sec:initial}) and we studied the charge exchange process as a function of the $P_s$ centre of mass velocity through the parameter $k_v$ defined in  equation \ref{eq:kv}. We are mostly interested in the values of $n_{P_s}$ in the interval 13-20, however we  performed the calculation for  $n_{Ps}$  spanning the range from 3  to 50.

As anticipated in the introduction, it turns out that  over the whole range of  $n_{P_s}$ values that we have investigated the cross section scales as $n_{P_s}^4$ and  $\sigma/n_{P_s}^4$ shows a universal behaviour as a function of $k_v$ as  figure \ref{fig:AllB0} shows. 

For $k_v \geq  2-3$   the scaled cross section $\sigma/n_{P_s}^4$ rapidly drops while 
$k_v \simeq 0.3$ is a threshold below which it raises as $\dfrac{1} {k_v^2}$ and it reaches interesting  high values. 
The right plot of figure \ref{fig:AllB0} shows the region of low $k_v$ values and a fit with the function $ \sigma/n_{P_s}^4 [cm^2]= \dfrac{s_1}{k_v^2} + s_2 $.

The same points plotted  as a function of the centre of mass energy of  positronium $E_{Ps}^{cm}$ are shown in figure \ref{fig:AllB0Ene}. The $\dfrac{1} {k_v^2}$ law of course translates to an increase of the cross section as $1/E_{Ps}^{cm}$.
This trend is the same already found using  the two-centre convergent close-coupling (CCC)  method for  $ n_{P_s}=2, 3 $  \cite{Charlton1}, \cite{Charlton2} and,
according to our knowledge, this is the first time that this result is shown for collisions involving Rydberg positronium and antiprotons.

The onset of the  $1/ E_{Ps}^{cm}$ regime approximately scales as $1/n_{Ps}$.

The rise of the cross section in case of  low energy  Rydberg positronium and its  high values are two results 
of extreme interest for the design and the optimisation of the antihydrogen experiments.

\begin{figure}
\begin{center}
\hspace{-5 mm}
\centering{\includegraphics[width = 0.5\textwidth]{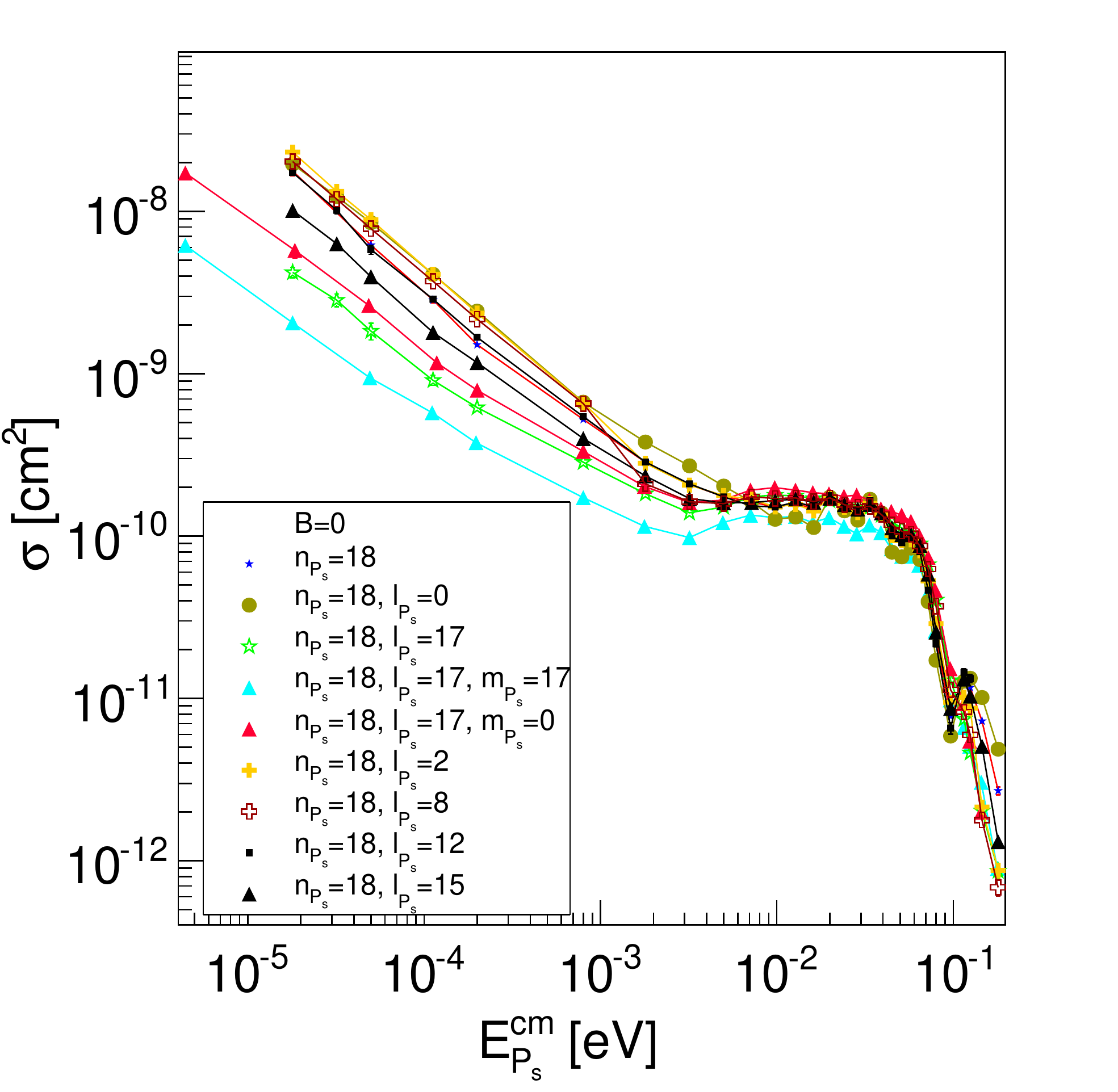}}
\caption{Charge exchange cross section as a function of the Ps center of mass energy for $n_{Ps}=18$ and various values of the angular momentum quantum numbers.
The plot with  $l_{P_s}$ or $m_{P_s}$  not specified has been obtained with a statistical distribution of angular momenta. The lines simply connect the points to help the  graphical interpretation.}
\label{fig:AllB0nPs18Ene}
\end{center}
\end{figure}  

\begin{figure}
\begin{center}
\hspace{-5 mm}
\centering{\includegraphics[width = 0.5 \textwidth]{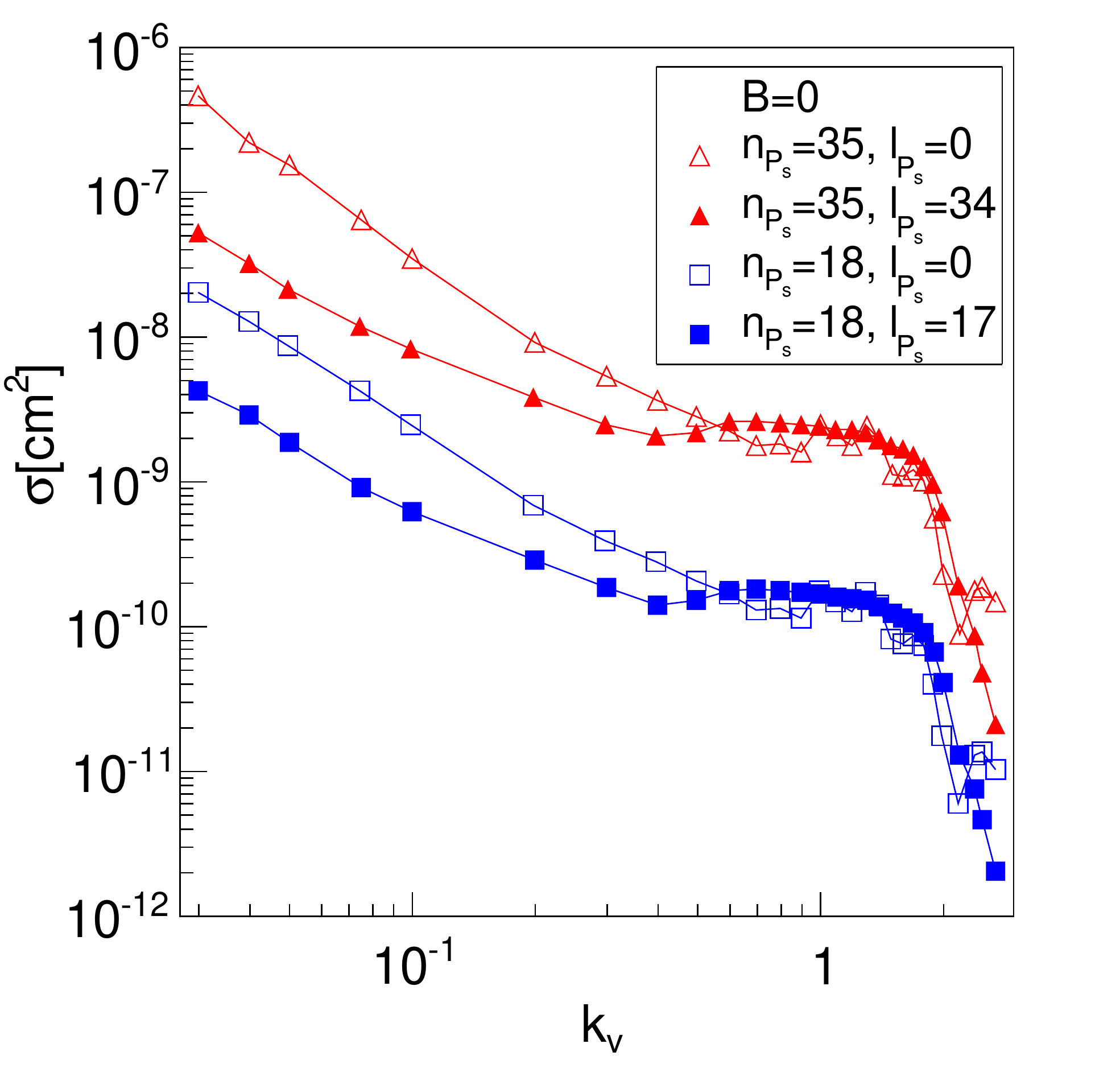}}
\caption{Charge exchange cross section for positronium in the initial state with $n_{Ps}$=18 and 35 and with the extreme values of the angular momentum quantum number
($l_{P_s}=0$ and $l_{P_s}=n_{P_s}-1$) as a function of $k_v$.  The lines simply connect the points to help the  graphical interpretation. }
\label{fig:Variousl}
\end{center}
\end{figure}    

\begin{figure}
\begin{center}
\centering{\includegraphics[width = 0.45\textwidth]{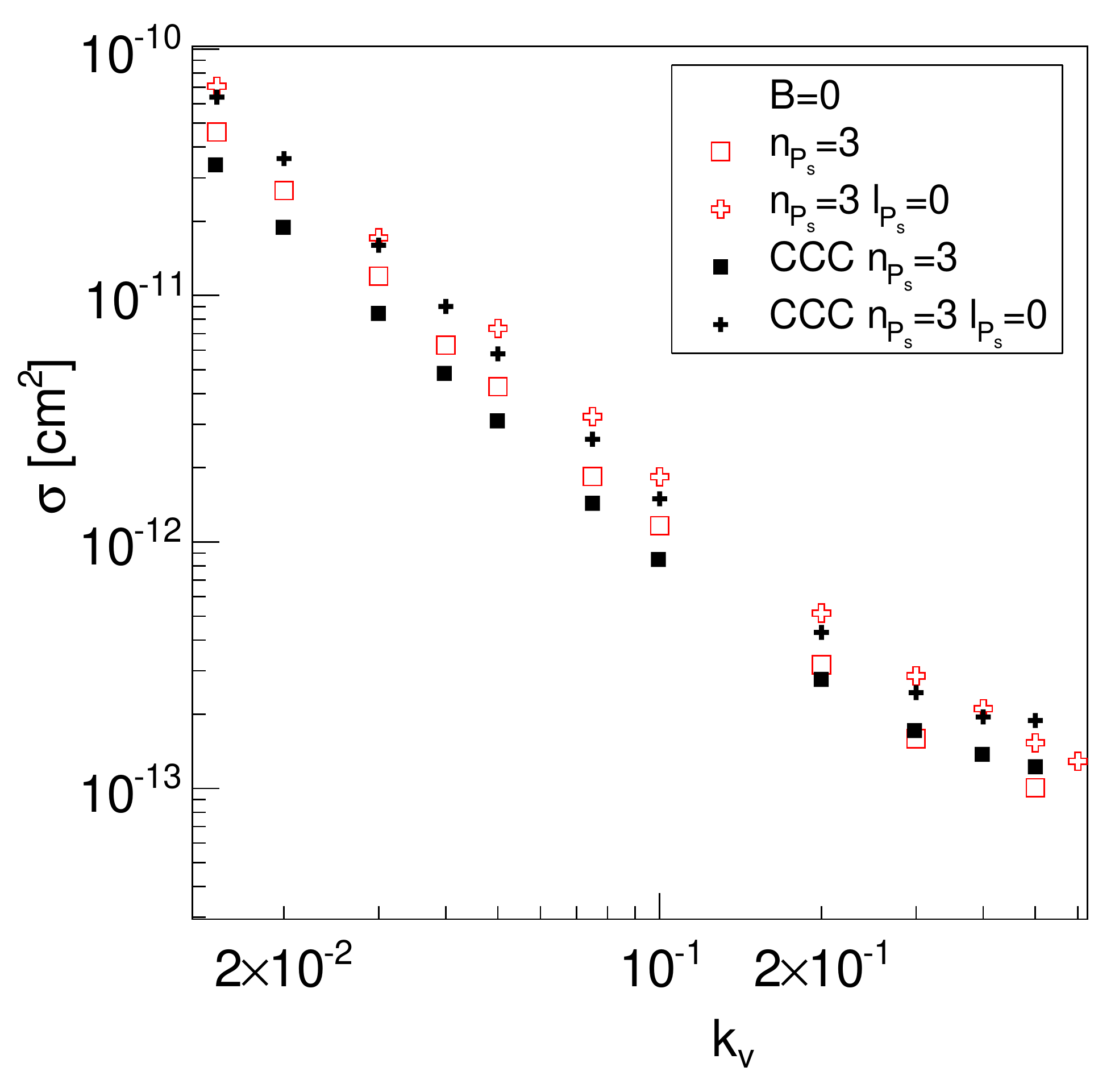}}
\caption{Low energy charge exchange cross section calculated with our CTMC method (open red squares and open red crosses) and with the CCC method (filled black squares and filled black crosses) of \cite{Charlton1}, \cite{Charlton2}.}
\label{fig:CCC}
\end{center} 
\end{figure}    

Generally for a fixed value of  $n_{Ps}$  the cross section depends on the initial angular state of the positronium: this is exemplified in   figure \ref{fig:AllB0nPs18Ene}
 for $n_{Ps}=18$.
Particularly the differences are enhanced in the low energy region ($1/ E_{Ps}^{cm}$ regime) being   the charge exchange probability significantly higher for the lowest angular momentum states than for the highest ones.
This general tendency is reproduced for other values of $n_{Ps}$. Figure \ref{fig:Variousl} shows the cross section as a function of the reduced velocity $k_v$ for $n_{Ps}=18$ and
$n_{Ps}=35$ and the extreme values of the angular momentum ($l_{P_s}$=0 and  $l_{P_s}=n_{P_s}$-1).  The $k_v$ threshold below which the cross section approximately scales as $E_{Ps}^{-1}$ is
about 0.9 for l=0 and about 0.4 for l=n-1. 
We have also investigated for $l_{P_s}=n_{P_s}-1$ the role of $m_{P_s}$ and  found  that in the low energy region there is also a dependence of the cross section upon $m_{P_s}$ with
high $m_{P_s}$ giving a lower cross section. Examples are in figure \ref{fig:AllB0nPs18Ene}.

Our results are in perfect agreement with the CTMC calculation reported in \cite{Lu} for $n_{P_s}$=50 and limited to $k_v>0.5 $.

The accuracy of the classical CTMC is expected to increase with the values of the principal quantum number of $P_s$ but the  limits of the validity of the classical approach are unclear. The comparison between the  low velocity cross section  obtained with the CTMC  and the result of the CCC method described in \cite{Charlton1} \cite{Charlton2}  for  collisions involving $n_{P_s}=3$  and $l_{P_s}=0$  or  $n_{P_s}=3$ and a statistical distribution of $l_{P_s}$   is reported in figure
\ref{fig:CCC}. The two methods show the same shape of the cross section  as a function of the collision velocity with  discrepancies in the numerical values of few ten $\%$.

  The CTMC and CCC methods also agree in describing  the qualitative proportion of the  $\bar H$ final state distribution with $n_{P_s}=3$ in the $1/E_{P_s}^{cm}$ regime: the dominant channel in the one originating $\bar H$ with $n=4$ followed by that giving $n=3$ with the production of antihydrogen with $n=1$ and $2$ accounting only for few percent or less of the total.
 However, as example,  in case of  collisions induced by $n_{P_s}=3$ and $l_{P_s}=0$, the ratio between the cross section for producing  $n_{\bar H}$ =4 and $n_{\bar H}$ =3 is close to 20 in CCC while it is  slightly higher than 10 in the CTMC approach. 

\subsection{Distribution of the antihydrogen quantum numbers}

\begin{figure*}
\begin{center}
\hspace{-5 mm}
\centering{\includegraphics[width = 1 \textwidth]{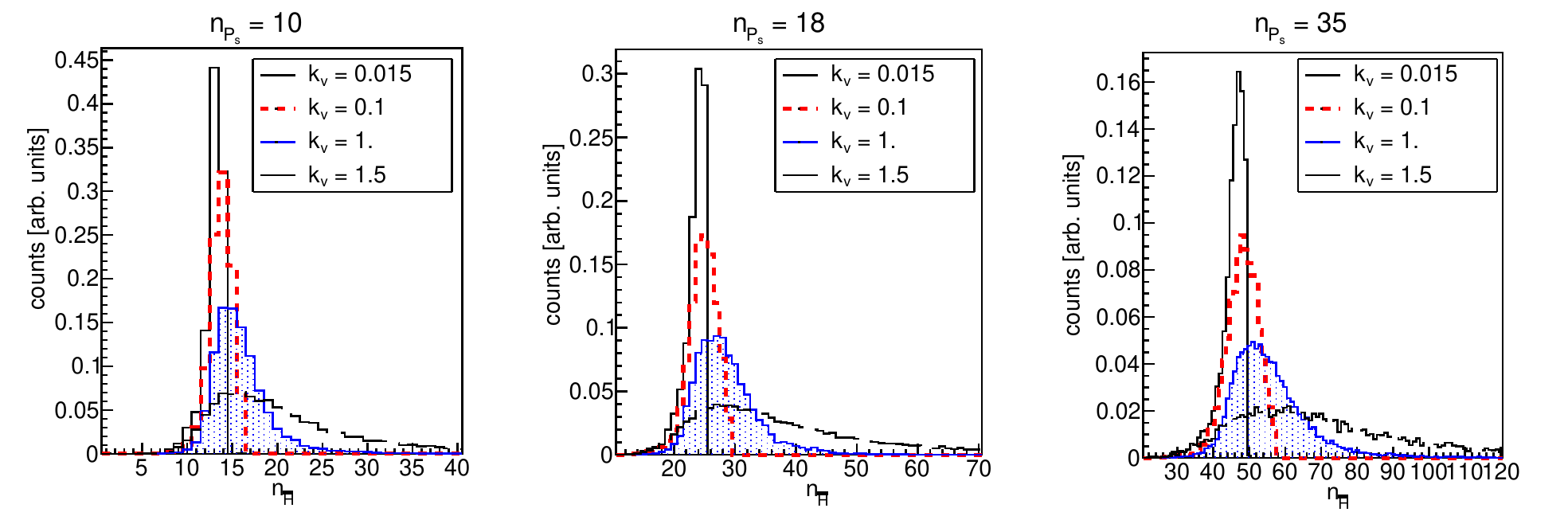}}
\caption{Normalised distribution of the principal quantum number of antihydrogen produced by interaction with $P_s$ with  $n_{Ps}$=10, 18, 35   and all the values of $l_{P_s}$ and  $m_{P_s}$. 
 B=0.
 }
\label{fig:Ndistrib}
\end{center}
\end{figure*}  

Our CTMC model shows that the antihydrogen atoms are always formed with  a distribution of the principal quantum number $n_{Hbar}$ even when the incoming $P_s^*$ has a fixed $n_{P_s}$.
The distribution is roughly peaked around $n_0 =\sqrt{2} n_{Ps}$ corresponding to the same binding energy of the positron in the initial positronium and in the final antihydrogen. 
From standard  kinematic arguments it follows that the antihydrogen formation in the limit of both  positronium and antiproton at rest can only happen if the Q value of the reaction, that is the difference of the binding energy of the initial positronium and the final antihydrogen,   is positive
\begin{equation}
Q= \frac{1}{4 n_{P_s}^2} - \frac{1}{2 n_{\bar H }^2}
\label{eq:Qvalue}
\end{equation}
The condition $Q>0$ translates into $n_{\bar H} \leq \sqrt{2} n_{P_s}$. The results of the CTMC consistently show that in the low velocity regime,
corresponding to the $1/ E_{Ps}^{cm}$ scaling, the distribution of  the principal quantum number of the formed antihydrogen has a small spread,  is asymmetric, peaked around $n_0$ with a population of antihydrogen with $n_{\bar H} > n_0$  negligible. 
The CTMC also shows that   when $k_v$ is in the range (0.3,1) $n_{Hbar} $ are produced with a bell shaped  distribution  peaked at $n_0$ and with a FWHM $\simeq 0.3$ $ n_0$. For larger values of
$k_v$ the produced antihydrogen has a wider distribution of principal quantum numbers with tails extending up to several $n_0$.
Figure \ref{fig:Ndistrib} shows  an example of the above mentioned  effect for  $n_{Ps}=10, 18, 35$. The shape of the distribution is practically independent upon $n_{P_s}$ when plotted as a function of $\dfrac {n_{\bar H}} {n_0}$.

\subsection{Angular momentum of the antihydrogen atoms}

\begin{figure}
\begin{center}
\centering{\includegraphics[width = 0.48 \textwidth]{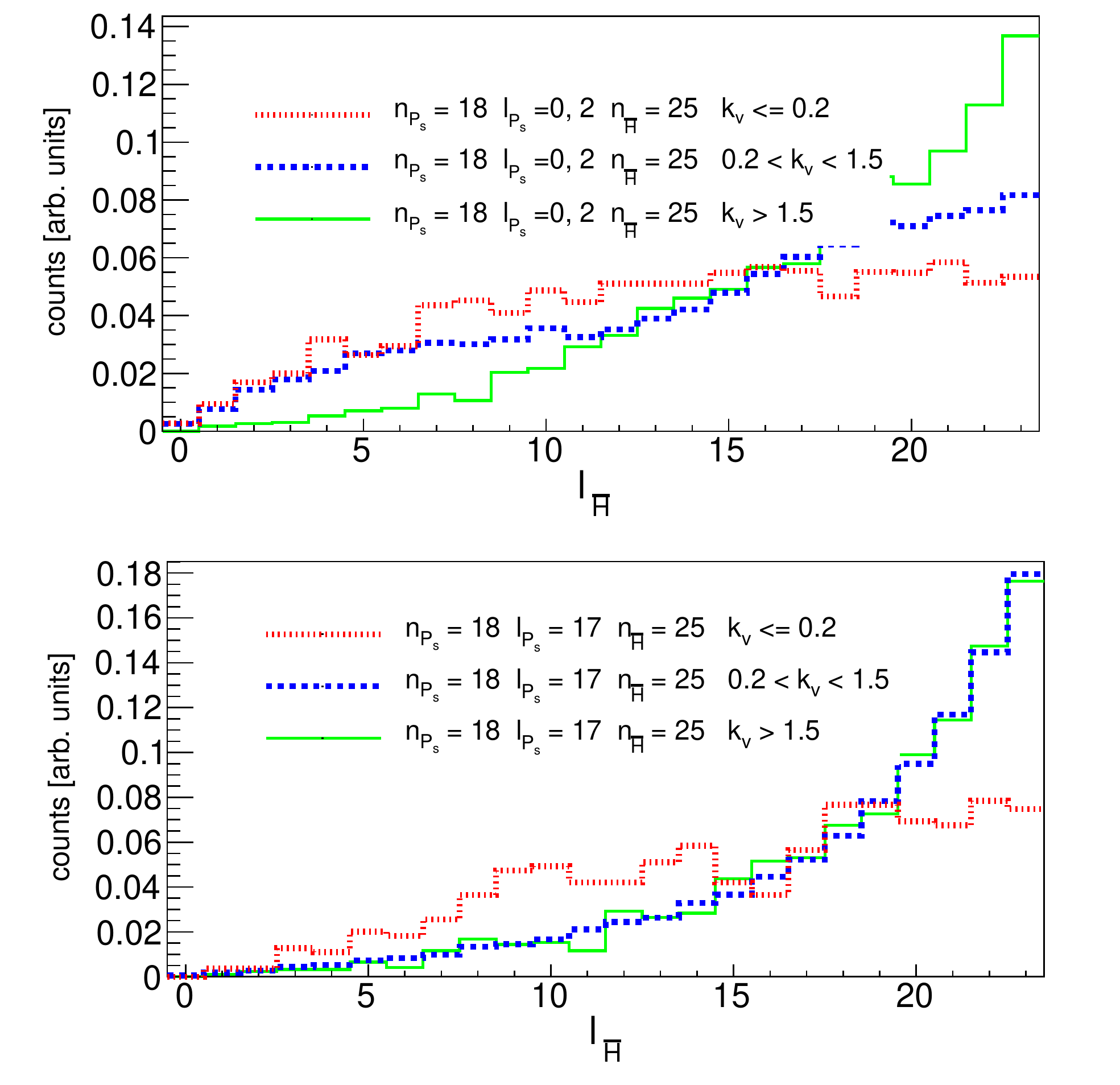}}
\caption{Distributions (normalised to unit area)  of $l_{\bar H}$ obtained by collisions with positronium in selected states of angular momentum $l_{P_s}$. The top panel refers to $l_{P_s}$=0 and 
$l_{P_s}$=2 and the bottom one to $l_{P_s}$=17. The three curves show results obtained selecting different intervals of the $k_v$ parameter as shown in the legend. $k_v <0.2$ is roughly the onset of the $\dfrac {1¹}{E_{P_s}^{cm}}$ regime.}
\label{fig:lDistrib}
\end{center}
\end{figure}  

Antihydrogen atoms are always  produced with a wide distribution of angular momenta $l_{\bar H}$ and 
for each of them  all the states with all possible values of $m_{\bar H}$ are statistically populated. 
The shape of the distribution of $l_{\bar H}$ shows features that depend on $k_v$ and also on the initial $l_{P_s}$. 
An example of the above effect is shown in figure \ref{fig:lDistrib}, where antihydrogen formation is studied for $n_{P_s} = 18$  for extremal values of $l_{P_s}$.
In both cases in the low velocity regime $k_v <0.2$ antihydrogen is produced with an angular momentum distribution  that does not rise significantly as a function of $l_{\bar H}$,
while for increasing values of $k_v$ high values of angular momentum become more probable. Note that in case of equal probability for each angular momentum the $l_{\bar H}$
distribution should show a linear shape.

\subsection{Velocity of the antihydrogen}
As discussed in the introduction, the velocity of the antihydrogen is an important parameter that influences the possibility to perform precision experiments.
Here we are assuming that the antiproton is initially at rest.
If this condition is not fulfilled  the recoil velocity here calculated has to be properly added to the initial antiproton velocity.

\begin{figure}
\begin{center}
\centering{\includegraphics[width = 0.45\textwidth]{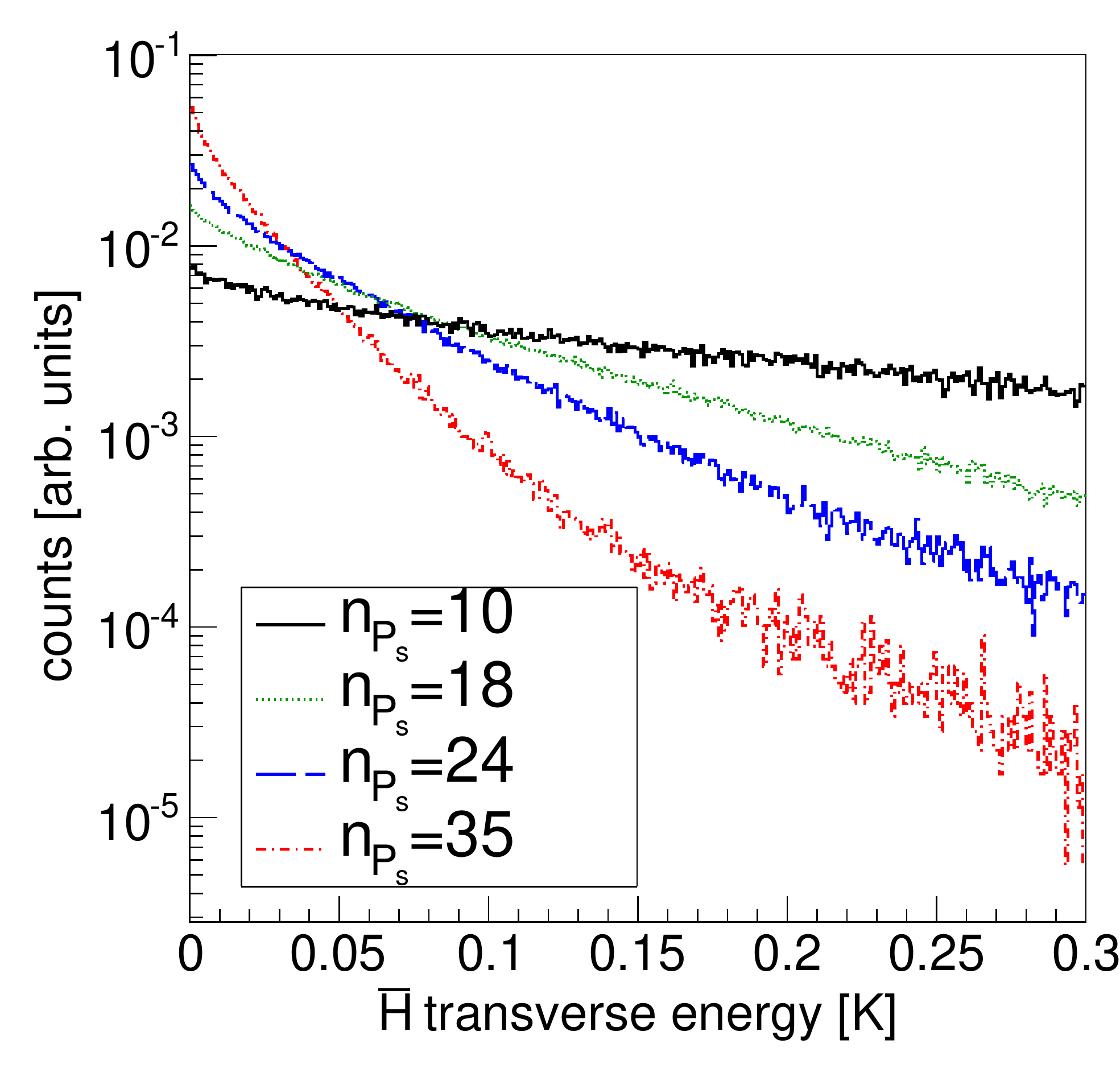}}
\caption{Distributions (normalised to unit area)  of the recoil  kinetic energy (expressed in Kelvin) of the antihydrogen in the direction  transverse to z 
for four different values of $n_{P_s}$. The interaction of antiprotons with high Rydberg states of $P_s$ produces colder antihydrogen.
 B=0. }
\label{fig:RadialEnergy}
\end{center}
\end{figure}

The recoil velocity of the antihydrogen 
in  the direction perpendicular to the flight direction of the positronium (assumed as z) has a null mean value and a spread that decreases while $n_{P_s}$ increases. For a fixed $n_{P_s}$ it does not significantly depends on $k_v$.  The fraction of antihydrogen with low radial recoil energy produced  by interaction of antiprotons with fixed  $n_{P_s}$ positronium 
 increases with  $n_{P_s}$ as  figure 
\ref{fig:RadialEnergy} shows thus indicating that high Rydberg states of positronium are preferred if one aims to cold antihydrogen.

\begin{figure}
\begin{center}
\centering{\includegraphics[width = 0.45\textwidth]{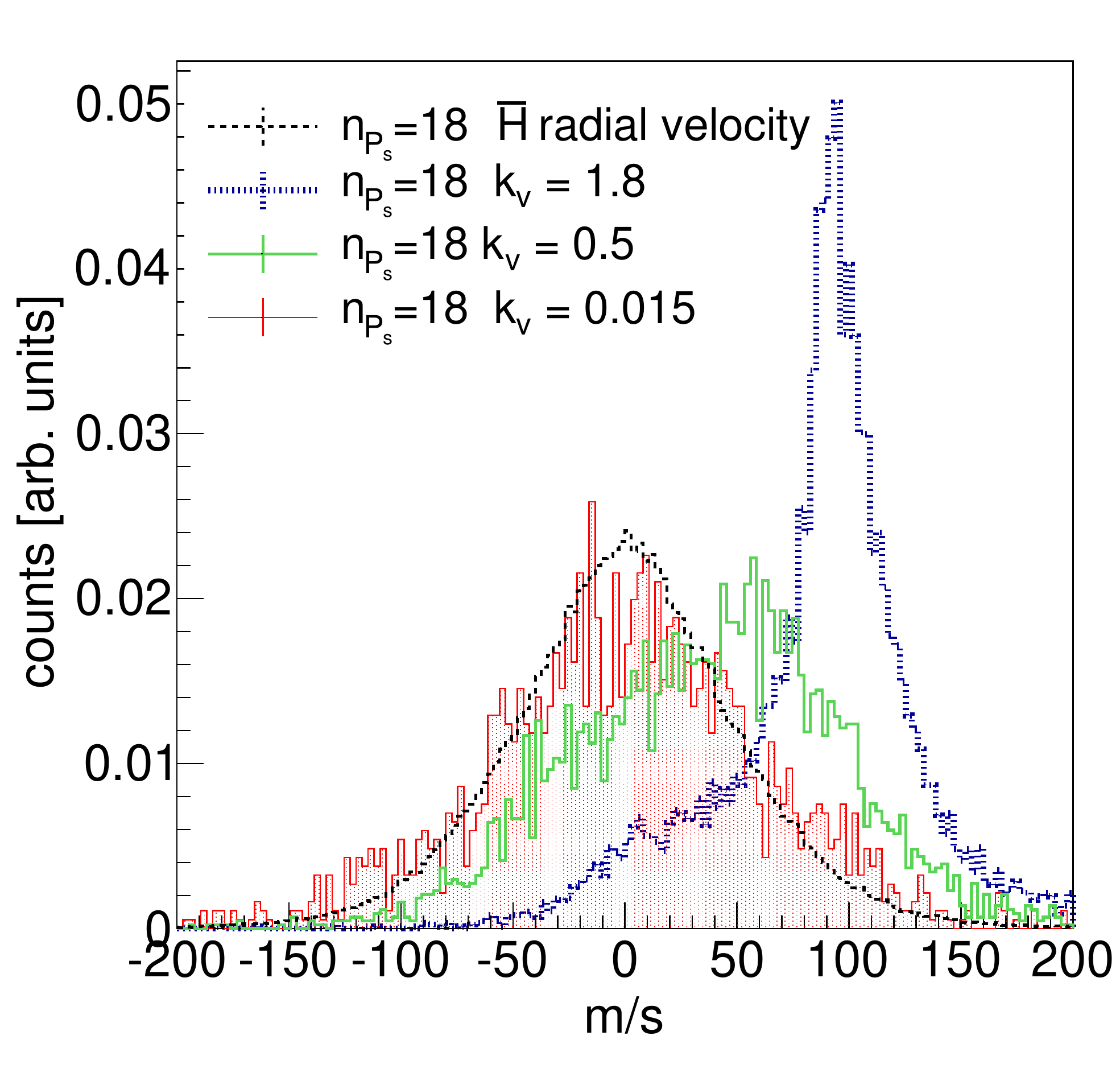}}
\caption{ Distribution of the velocity of the antihydrogen in the flight direction of the positronium (z) obtained with $n_{P_s}$= 18 and some selected values of $k_v$.
For comparison the distribution of the velocity in one of the transverse direction is reported. All the histograms are normalised to unit area.
 B=0.
 }
\label{fig:AxialVelocity}
\end{center}
\end{figure}

The antihydrogen gets a small boost  (as reported in figure  \ref{fig:AxialVelocity}) along the flight direction of the incoming positronium related to its centre of mass velocity. 
This effect is particularly interesting if one is aiming to form a beam of cold antihydrogen and it is required that positronium fly toward the antiprotons along the wished beam direction. However it should be observed that the antihydrogen boost is significant only when $k_v$ is above  the $1/k_v^2$ regime and then  a proper tradeoff between flux of produced antihydrogen and its directionality   has to be practically considered.

\subsection{The impact parameter}
\begin{figure}
\begin{center}
\centering{\includegraphics[width = 0.45 \textwidth]{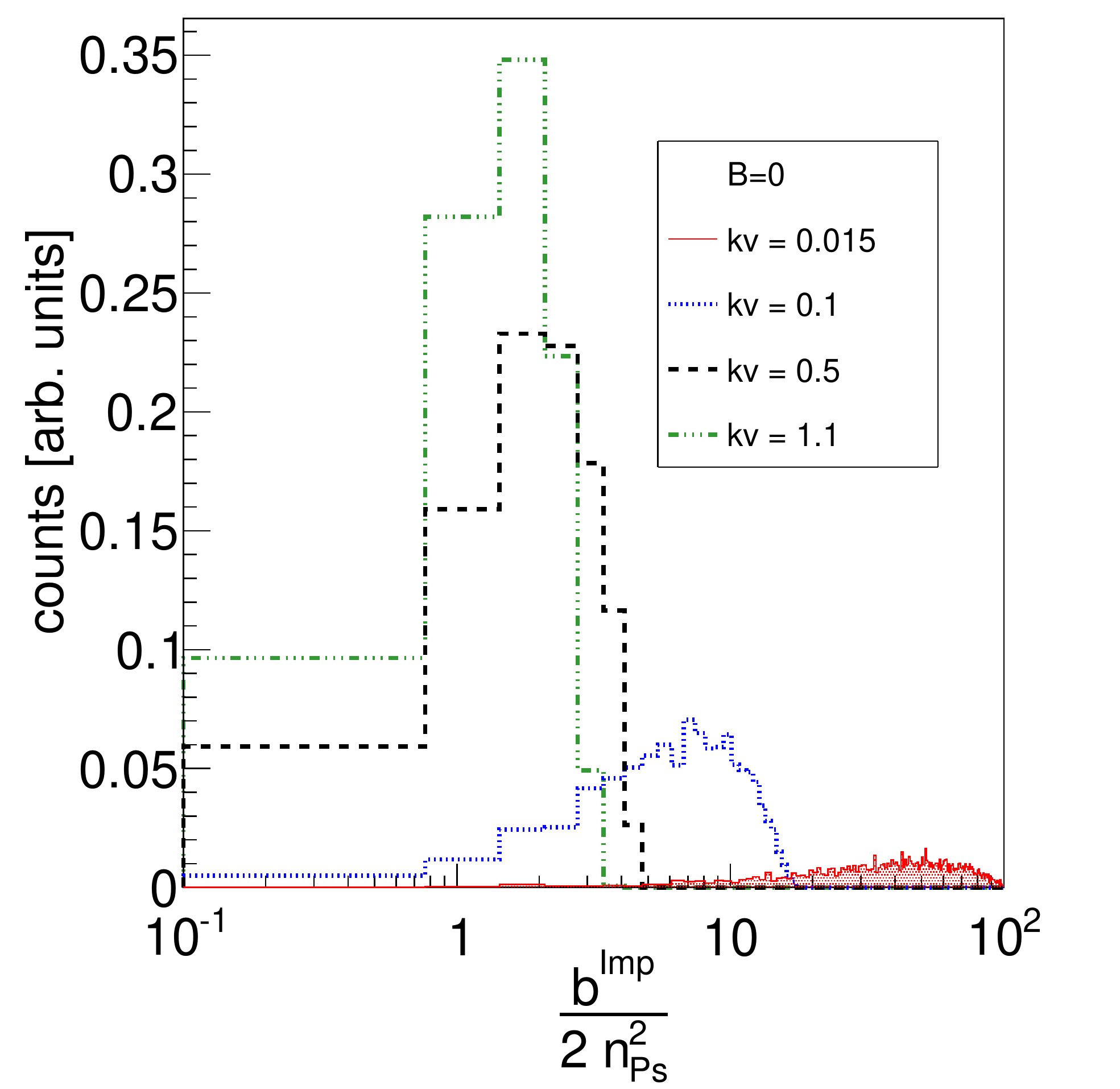}}
\caption{Distributions (normalised to unit area) of the impact parameters (scaled by the positronium orbit semi-axis major) of the collisions resulting in antihydrogen formation. B=0.}
\label{fig:Impact}
\end{center}
\end{figure}

The distributions of the impact parameter  of the collisions resulting in antihydrogen formation is shown in figure \ref{fig:Impact}.
The impact parameter is normalised to the size of the semi axis major of the positronium orbit and the distributions  are normalised to  unit area.
The shape of these scaled distributions is basically the same for all the values of the $n_{P_s}$ investigated.
Not surprisingly large impact parameters allow antihydrogen production only for low velocity collisions thanks to the relatively long time spent by the positronium in proximity of the antiproton.

\section{Charge Exchange in presence of magnetic field}
\subsection{Coupling between centre of mass and internal motion}

\begin{figure*}
\begin{center}
\centering{\includegraphics[width = 1.0 \textwidth]{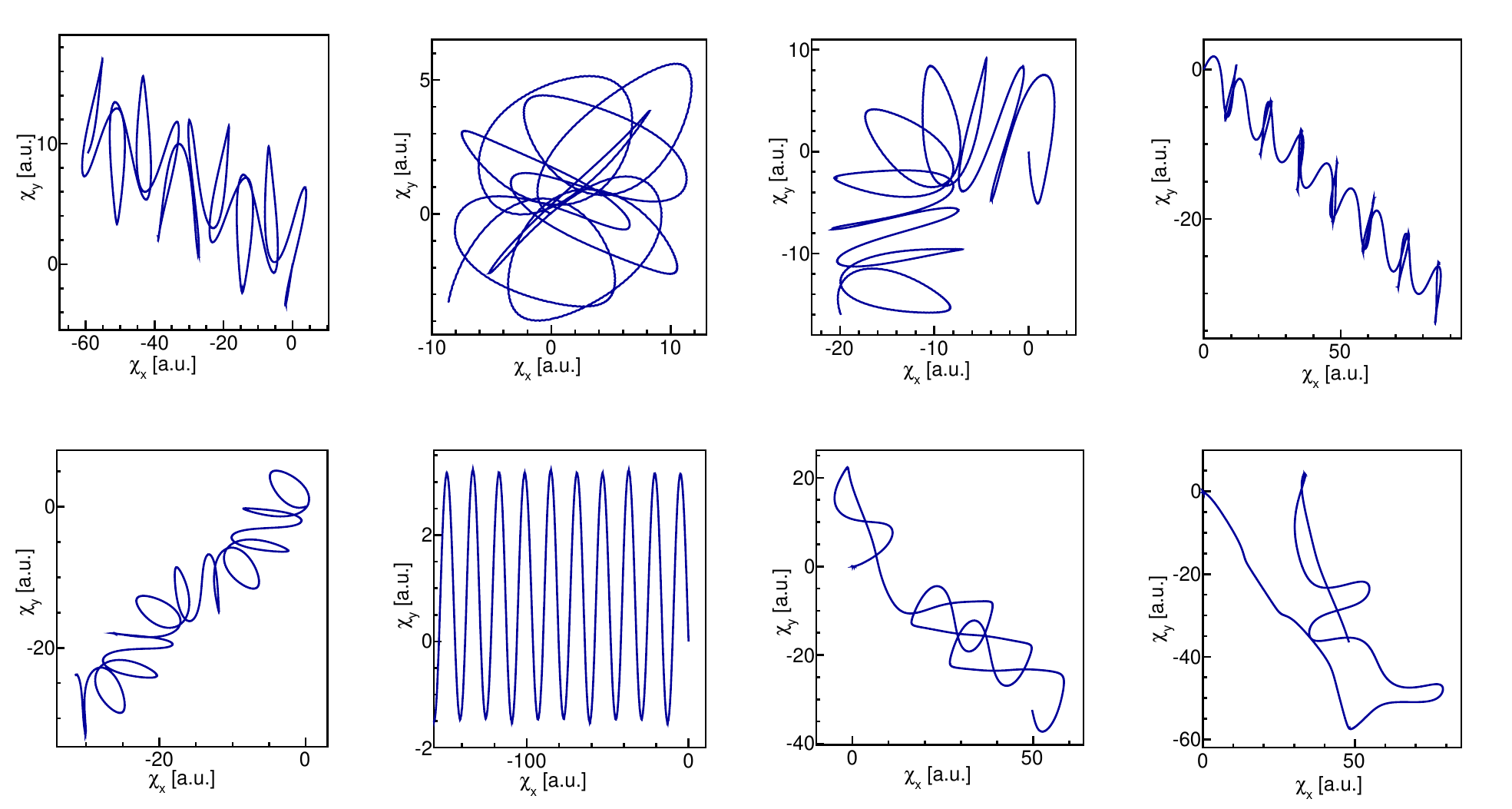}}
\caption{Example of projection in the $xy$ plane of some trajectories of the centre of mass of $P_s$ with $n_{P_s} =18 $ flying in 1 T magnetic field directed along $z$ for a time interval of 5000 $\tau_n$ following the adiabatic switching of the magnetic field. These trajectories have been calculated without  interaction with the antiproton. The  $P_{s}$ centre of mass is placed  in  (x=0, y=0, z=0) at the end of the adiabatic switching of $\vec B$.  A random elliptical trajectory is initially  selected as discussed in section \ref{sec:initial} and  the adiabatic switching procedure is then
performed raising the magnetic field in 1000 $\tau_n$. The shape of this $xy$ projections does not depend on the conserved $z$ velocity of the $P_s$ centre of mass. We plot the quantity $\chi_x= x^{cm}_{P_s}/(2 n_{P_s}^2)$ and $\chi_y= y^{cm}_{P_s}/(2 n_{P_s}^2)$ that is the transverse coordinates scaled by the size of the semi-axis major of the unperturbed elliptical trajectory of $P_s$.}
\label{fig:traject}
\end{center}
\end{figure*}  
We have extended the CTMC approach including  the presence of an external magnetic field $\vec B$. We consider here fields of moderate values
($B \simeq $ 0.5 -2  T) as foreseen in the AEgIS experiment  \cite{AEgIS}.

The magnetic field  influences the dynamics of the collision and  the initial and final  status of $P_s^*$ and $\bar H$.
Particularly important is the fact that 
the Hamiltonian of  a two-body charged system in magnetic field cannot be separated as the sum of two contributions one describing the centre of mass
and the other  one the internal motion as in the free field case. This result applies both to the description of positronium and antihydrogen;
nevertheless the separation is a good approximation only  in the limit of infinite mass of one of two particles. We do not then discuss this coupling between degrees of freedom for  antihydrogen
while we fully takes it into account for positronium.
The Hamiltonian of $P_s$ in presence of magnetic field is
\begin{equation}
H= \frac{1}{2} \left[\vec p_{e^+} + \vec A(\vec r_{e^+} \right ]^2 + \frac{1}{2} \left [\vec p_{e^-} - \vec A(\vec r_{e^-}\right]^2 - \frac{1}{r_{e^+e^-}}
\label{eq:Hfield}
\end{equation}
where the canonical momentum $\vec p_{e{^+}}$ is related to the mechanical momentum 
$\vec \pi_{e_{^+}}$ through the usual relation
 $\vec \pi_{e_{^+}} = \vec p_{e_{^+}} - \vec A(\vec r_{e^+})$  and $\vec \pi_{e_{^-}} = \vec p_{e_{^-}} +\vec A(\vec r_{e^-})$.
 
$\vec A (\vec r_{e^+,e^-})= \frac{1}{2} \vec B \wedge \vec r_{e^+,e^-}$ is the vector potential. 

It is useful to introduce the pseudo-momentum \cite{bib:PseudoMom} of the positron $\vec k_{e^+} = \vec p_{e^+} - \frac{1}{2} \vec B \wedge \vec r_{e^+}$ and 
of the electron $\vec k_{e^-} = \vec p_{e^-} +\frac{1}{2} \vec B \wedge \vec r_{e^-}$.
In absence of magnetic field the total mechanical momentum is conserved but in presence of magnetic field the total canonical momentum 
$\vec P_{P_s} = \vec p_{e_{^+}} +\vec p_{e_{^-}}$ does not commute with the Hamiltonian and it is not conserved. 
However  the total
pseudo-momentum $\vec K_{P_s}$ is conserved
\begin{equation}
\vec K_{P_s} = \vec k_{e^+}+ \vec k_{e^-} = \vec P_{P_s} +\frac{1}{2} \vec B \wedge (\vec r_{e^-}-\vec r_{e^+}) 
\end{equation}
Using  the centre of mass coordinate $\vec R_{P_s}$ and  pseudo-momentum $\vec K_{P_s}$ as  one set of canonically coniugated variables and the relative coordinates and momentum  $\vec r_{e^+e^-}$ and $\vec p_{e^+e^-}$ as second set 
 the Hamiltonian becomes
 \begin{equation}
H  = \dfrac{K_{P_s}^2}{4} -\dfrac{1}{2} (\vec K_{P_s}\wedge \vec B) \cdot \vec r_{e^+e^-}  + p_{e^+e^-}^2 + \dfrac{1}{4} (\vec B \wedge \vec r_{e^+e^-})^2 - \frac{1}{r_{e^+e^-}}
\label{eq:Hpseudo}
\end{equation}
 and the motion equation are then written  in a  form that clearly shows the coupling between internal and centre of mass degrees of freedom \cite{bib:PsMotion},\cite{bib:Pohl} 
 
\begin{equation}
\left \{ \begin{array}{lcccccccl}
\dfrac {d \vec R_{P_s}} {dt} = \dfrac{1}{2} \vec K_{P_s} - \dfrac {1}{2}  (\vec B  \wedge \vec r_{e^+e^-})\\
\\
\dfrac {d \vec K_{P_s}}{dt} =0 \\
\\
\dfrac{d\vec r_{e^+e^-}}{dt} = 2   {\vec p_{e^+e^-}}\\
\\
\dfrac{d \vec p_{e^+e^-}}{dt} = -\dfrac{1}{2} (\vec B \wedge \vec K_{P_s})  + \dfrac{1}{2} \vec B \wedge (\vec B \wedge \vec r_{e^+e^-}) - \dfrac{ \vec r_{e^+e^-}} { r_{e^+e^-}^3}    
\end{array}  \right.
\label{eq:Psmotion}
\end{equation}

One of the consequences of the internal and centre of mass motion coupling  coupling is that the centre of mass does not  move on  straight line trajectory as in the field free case. The centre of mass trajectory is related to  the time dependent  relative coordinate $\vec r_{e^+e^-}$ while the internal motion depends on the centre of mass through the conserved quantity $\vec K_{P_s}$.
These features  have been discussed in \cite{bib:PsMotion}, \cite{bib:Pohl} and \cite{bib:HinB}  where it is also underlined that the dynamics is not determined by  the energy $E$ and magnetic field strength
separately but only on the scaled quantity $\epsilon = E B ^{-2/3}$.  Varying $\epsilon$ from -3 to -0.1 the internal motion  undergoes a transition from regular motion to chaos.
With the magnetic fields and energies  here considered we expect to be in the fully regular regime.

\subsection{Construction of classical trajectories corresponding to quantum states of $P_s^*$ in magnetic field}

\begin{figure}
\begin{center}
\centering{\includegraphics[width = 0.4 \textwidth]{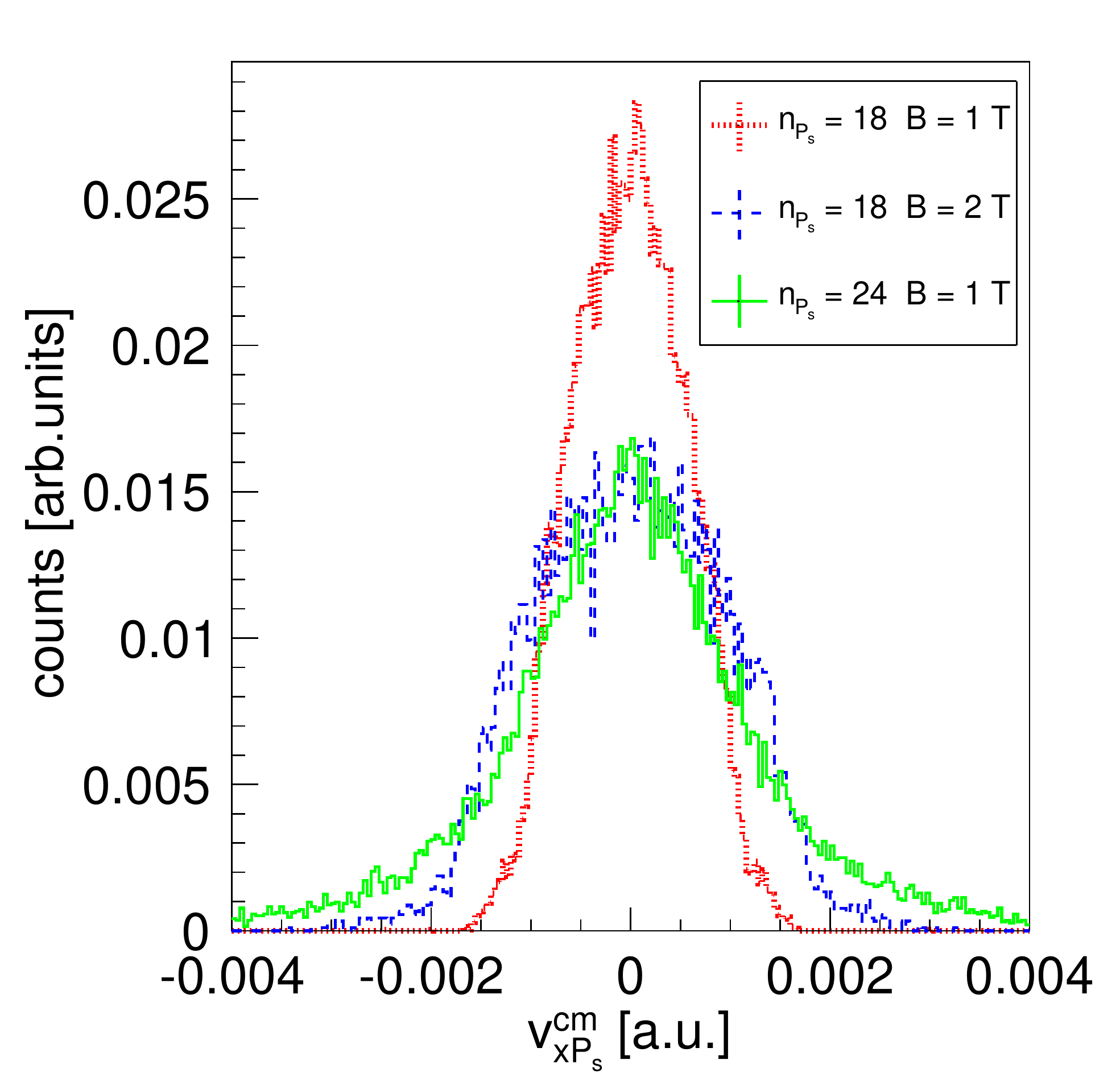}}
\caption{Example of distributions (normalised to unit area) of one component of the $P_s$ centre of mass transverse velocity  obtained after the adiabatic switching and randomisation procedure with $\vec B$ along the z axis. The initial transverse velocity is null. }
\label{fig:nPsTransverseVelocity}
\end{center}
\end{figure}  

\begin{figure}
\begin{center}
\centering{\includegraphics[width = 0.5 \textwidth]{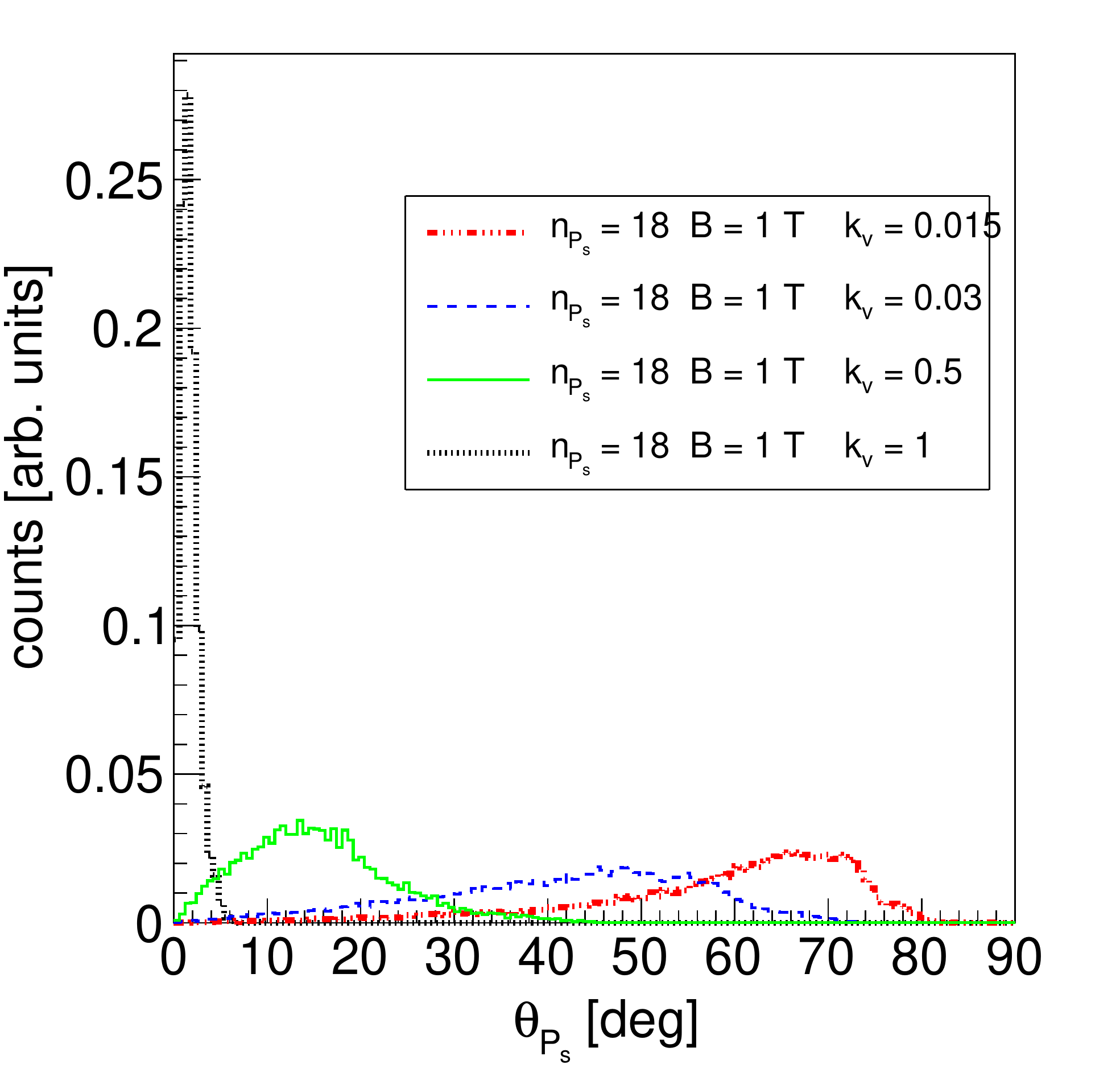}}
\caption{Distributions (normalised to unit area) of the angle $\theta_{P_s}$ between the axial and transverse centre of mass velocity of $P_s$ obtained after the adiabatic switching and randomisation procedure with B=1 T and $n_{P_s}$ =18. }
\label{fig:nPsAngle}
\end{center}
\end{figure}

We employed the adiabatic switching procedure \cite{AdiabSw} to construct trajectories corresponding to quantum states  of Rydberg $P_s$ in magnetic field.
This method is largely used for non-separable systems \cite{Adiab2} and it has been recently suggested for the  description of quantum states of hydrogen in magnetic field
\cite{Solovev}. An alternative approach 
is based on the modification of the classical elliptical trajectories
in    presence of magnetic field  as reported \cite{Bradenbrink}. This method, introduced for Rydberg atoms,  is not appropriate for Rydberg $P_s$ because it does not consider the coupling between center of mass and internal motion.
The adiabatic switching procedure automatically takes into account this coupling and produces a final state in which the center of mass and the internal degrees of freedom are coherently described.

We randomly selected  elliptical trajectories of $P_s^*$ in absence of magnetic field as described in section \ref{sec:initial} and we then followed the full motion (centre of mass and internal motion) of $P_s^*$
while the external magnetic field is adiabatically switched on. In practice we solved the motion equation for the Hamiltonian $H(t)$ (as in relations \ref{eq:Hfield}, \ref{eq:Hpseudo}) with the addition of a time-dependent magnetic field $\vec B^{adiab}(t) = \lambda(t) \vec B$ slowly rising from 0 to the final value $\vec B$. We 
tuned $\lambda(t)$ in such a way that the full field is reached after some thousands of periods
of the unperturbed elliptical motion $\tau_n =4 \pi n_{P_s}^3$. 
We have checked that the results about the cross section are stable as a function of the time used  to ramp the  magnetic field.  

Sampling the initial state from a microcanonical ensemble simply ensures that all possible initial states are considered. Note that 
  nor the angular momentum nor its $z$ component are conserved in general conditions with not null $K_{P_s}$.

The conservation of $\vec K_{P_s}$ leads to the conservation of the component of the centre of mass velocity of $P_s$ in the direction of the magnetic field.
 Instead the components transverse to the field  are not conserved and, as result of  the coupling between centre of mass and internal motion,  at the end of the adiabatic switching of the magnetic field we obtain a centre of mass velocity in the direction transverse to the magnetic field despite of its eventually  null initial value.
The trajectories of the centre of mass of Rydberg $P_s$ in magnetic field are then characterised by significant excursions in the  plane perpendicular to $\vec B$
and they show substantial deviations from the field free straight lines.
Figure \ref{fig:traject} refers to  $n_{P_s}=18$, B=1 T directed along $z$ and it 
shows some arbitrary example of centre of mass trajectories of $P_s^*$ projected in the  $x$ $y$ plane. They have been obtained tracking the $P_s^*$ motion  without interaction with antiproton   for a time interval of 5000 $\tau_n$ after the adiabatic switching of the magnetic field.

Discussions  about the centre of mass trajectories in case of $\vec K_{P_s} =0$ can be found  in \cite{bib:PsMotion}; we did not attempt here to perform a classification or  a general study of the features of these trajectories in the general case of a  not null $\vec K_{P_s}$.

The internal motion is still described within  good approximation by elliptical trajectories with not constant semi-axis minor (corresponding to a not conserved angular momentum) and with not constant orientation in space (corresponding to a not conserved projection of the angular momentum in the direction of the magnetic field).
For particular values of $K_{P_s}$ and $B$ one would expect the existence of long lived  delocalized states of positronium as a minimum of the potential could appear
in addition  to the Coulomb singularity  at null inter-particle distance. These states are predicted to appear when  the transverse pseudo-momentum 
is above a critical value $K_c$ with $K_c= (27 B /2)^{1/3}$ \cite{bib:PsDelocalized}.
These delocalized states are the initial states in the calculation of the charge exchange cross section in magnetic field in \cite{Lu} but they do not play a role here.

In order to fully randomise the $P_s^*$ initial conditions to be used in the charge exchange process, after the completion of the adiabatic switching of the magnetic field, we followed the motion of $P_s^*$ without interaction with the antiproton for a random time interval of few thousand $\tau_n$.
We used the centre of mass velocity and  the position and velocity of the internal  motion obtained at the end of this randomisation procedure  as initial values of the full
three-body tracking in magnetic field with interaction with the antiproton. The choice of the initial position of the centre of mass of $P_s^*$ is discussed in section \ref{XSecB}.

Figure \ref{fig:nPsTransverseVelocity} shows  an example of the distributions of the centre of mass velocity along the $x$ direction obtained at the end of the adiabatic switching and randomisation procedures with $\vec B$ along $z$. A similar  shape is obtained for the $y$ component. 
Note that, depending on the $k_v$ value, the transverse centre of mass velocity acquired by the $P_s^*$   can be  a small fraction  of the axial one or it can be even larger than that.
In any case, the  non null radial velocity of the centre of mass  has the consequence that  $P_s^*$  is emerging from  a given position
 with an angle $\theta_{P_s}$ with respect to the $z$ axis. Figure \ref{fig:nPsAngle} shows some distributions of this angle in case of $n_{P_s}=18$ and $B_z=1$ T.

\begin{figure}
\begin{center}
\centering{\includegraphics[width = 0.5\textwidth]{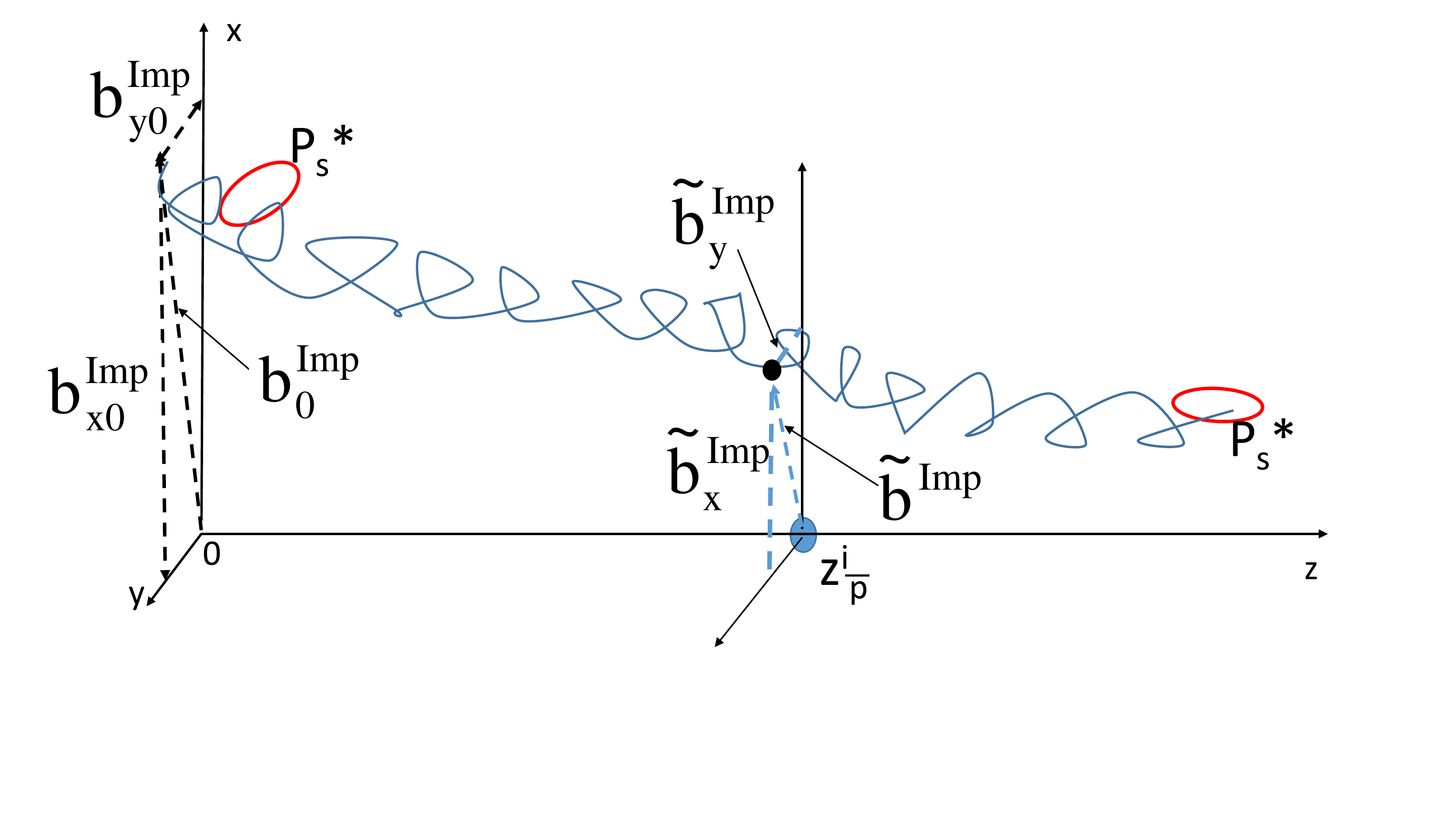}}
\caption{Definition of the geometry of the collision in presence of magnetic field.
The curve shows a pictorial view of a trajectory of  the $P_s^*$ centre of mass trajectory  without interaction with the antiproton.
The  impact parameter  used for the calculation of the cross section in presence of magnetic field is $\tilde b^{Imp}$ and it represents the distance betwen the $P_s^*$ trajectory and the 
point $z_{\bar p}^i$ evaluated in the plane $z = z_{\bar p}^i$.
$\tilde b^{Imp}= \sqrt{ (\tilde b_x^{Imp})^2 + (\tilde b_y^{Imp})^2}$.
Also shown in the figure are the radial coordinates   $(b_{x0}^{Imp},b_{y0}^{Imp})$ of the trajectory in $z=0$  
and, being the trajectory not a straight line,  
 they  differ from
$(\tilde b_x^{Imp},\tilde b_y^{Imp})$.  Without magnetic field $\tilde b_x^{Imp} = b_{x0}^{Imp}$  and $\tilde b_y^{Imp} = b_{y0}^{Imp}$.}
\label{fig:ImpactWithB}
\end{center}
\end{figure} 

\subsection{Impact parameter for collisions in presence of magnetic field}
\label{XSecB}
The definition of  cross section and impact parameter in presence of magnetic field deserves some caveats related to the curved trajectories of the $P_s^*$ center of mass in absence of interaction with  the antiproton.
Figure \ref{fig:ImpactNoB} shows the standard definition of the impact parameter $b^{imp}$: it is the distance that would be the distance of closest approach between the projectile
($P_s^*$) and the target ($\bar p$) in absence of interaction under the assumption that the unperturbed trajectory of the projectile would be a straight line \cite{Taylor}.
In equivalent way  we can draw a  line  parallel  to unperturbed projectile trajectory and passing trought the target center  (this is the $z$ axis in all this work) and  see that  $b^{Imp}$ is the distance between these two parallel lines.

When dealing with  collisions of  $P_s^*$ in magnetic field, we have adopted a definition of impact parameter 
$\tilde b^{Imp}$ 
that has the property $\tilde b^{Imp} \rightarrow b^{Imp}$ when $ B \rightarrow 0$ that is it reproduces  the standard definition for vanishing magnetic field.

Figure \ref{fig:ImpactWithB} shows a pictorial view of a 
the geometry of the collision in presence of magnetic field.
$\tilde b_{Imp}$ is defined as the distance between the $P_s^*$ center of mass trajectory in absence of interaction and the target antiproton evaluated in the plane $z= z_{\bar p}^i$. 
The figure \ref{fig:ImpactWithB} also shows   $b_0^{Imp}$ defined as the distance between the same  $P_s^*$ center of mass trajectory   and the $z$ axis evaluated
in the plane $z=0$.  Note that $\tilde b^{Imp}$ and $b_0^{Imp}$ are in general different.

The cross section results  shown below   have  been calculated according to equation \ref{eq:Xsec} using
 $\tilde b^{Imp}$ as impact parameter. This is  the same approach used in \cite{Lu}.
Precisely, we   uniformly   generated  within a circle
of radius $\tilde b_{max}^{Imp}$
 the points with coordinates  ($\tilde b_x^{Imp}, \tilde b_y^{Imp}$) in the plane $z=z_{\bar p}^i$.
 \begin{figure}
\begin{center}
\centering{\includegraphics[width = 0.5 \textwidth]{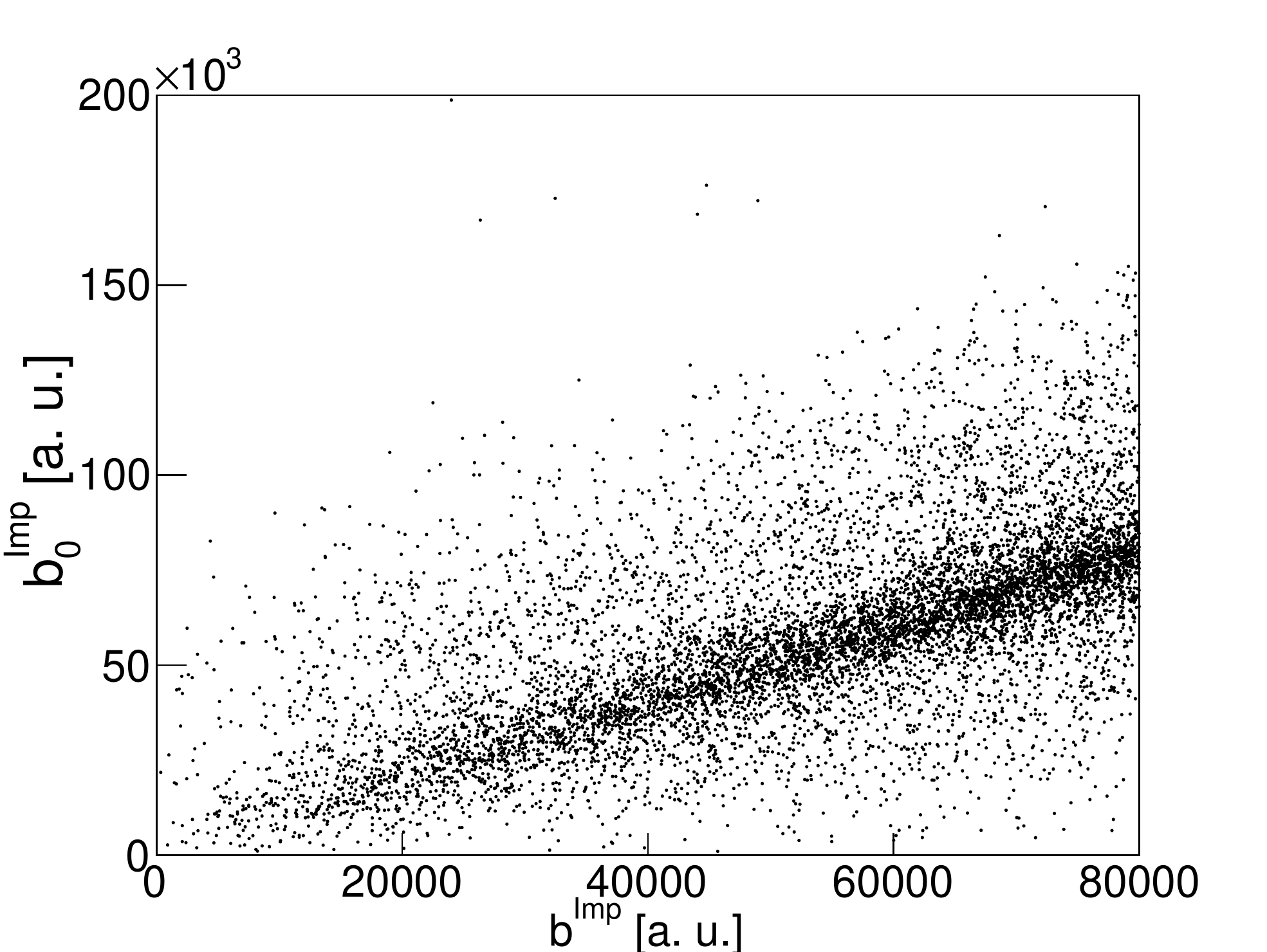}}
\caption{Scatter plot of $b_{0}^{Imp}$ versus $b^{Imp}$ obtained with $n_{P_s}=18$, B=1 T directed along $z$,
$k_v=0.015$ and method 2. The starting point of the $P_s$ centre of mass trajectory which is  back tracked until it reaches $z=0$  is uniformly generated within a circle with radius $b_{max}^{Imp} =8 \cdot 10^4$ AU}
\label{fig:ScatterPlot}
\end{center}
\end{figure}  
After having performed  the adiabatic switching of the magnetic field and the randomization  we have placed the $P_s^*$ in the point ($\tilde b_x^{Imp}, \tilde b_y^{Imp}, z_{\bar p}^i$). 
Then we have followed the $P_s^*$  full motion back in time 
(setting $v_z^{cm}<0$ and changing the sign of velocities and the direction of the magnetic field) 
without interaction with the antiproton until it reaches the position $z=0$. We call ($b_{0x}^{Imp}$ $b_{0y}^{Imp}$) the radial coordinates reached by the centre of mass 
when $z=0$.
Then  we inverted again  the sign of the velocity, we restored the initial direction of the magnetic field and we solved the three-body problem with the antiproton interaction switched on;   the positronium  starts from the position ($(b_{0x}^{Imp}, b_{0y}^{Imp}, 0)$) with the rest of the kinematic variables resulting from the back propagation procedure.

As example figure \ref{fig:ScatterPlot} shows all the values of $b_{0}^{Imp}$ obtained for each $b^{Imp}$ with $n_{P_s}=18$, B=1 T directed along $z$  and
$k_v=0.015$.

 



\subsection{Cross section results in  magnetic field}

\begin{figure}
\begin{center}
\centering{\includegraphics[width = 0.5 \textwidth]{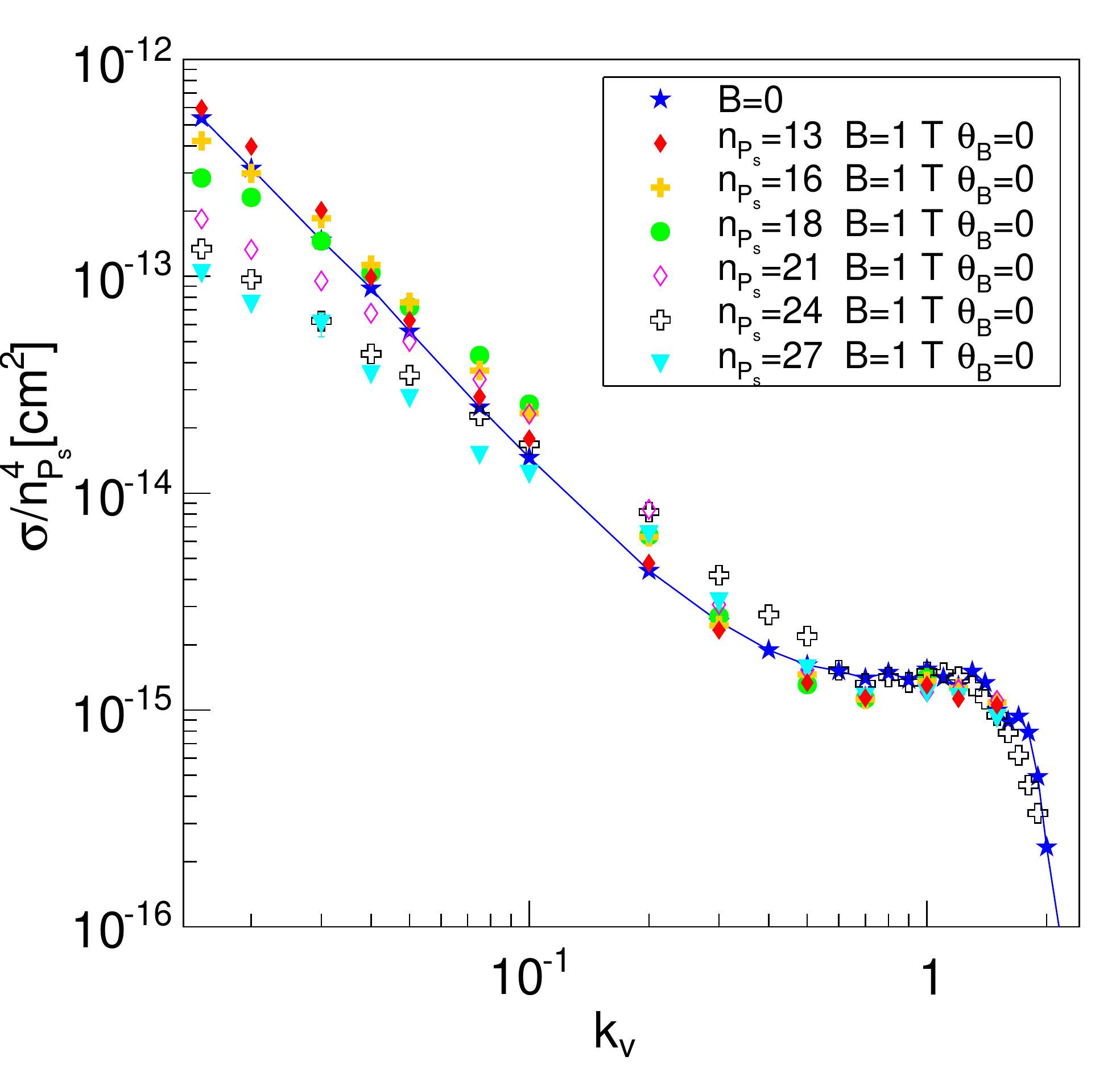}}
\caption{Charge exchange cross section divided for $n_{P_s}^4$ as a function of $k_v$ calculated with B=1 T, $\Theta_B=0$ and with  $n_{P_s}$=13, 16, 18,  21, 24, 27.
 For comparison the normalised zero magnetic field cross section is reported (blue stars). Compare with figure \ref{fig:AllB0}.}
\label{fig:B1TVariousn}
\end{center}
\end{figure}  

\begin{figure}
\begin{center}
\centering{\includegraphics[width = 0.5 \textwidth]{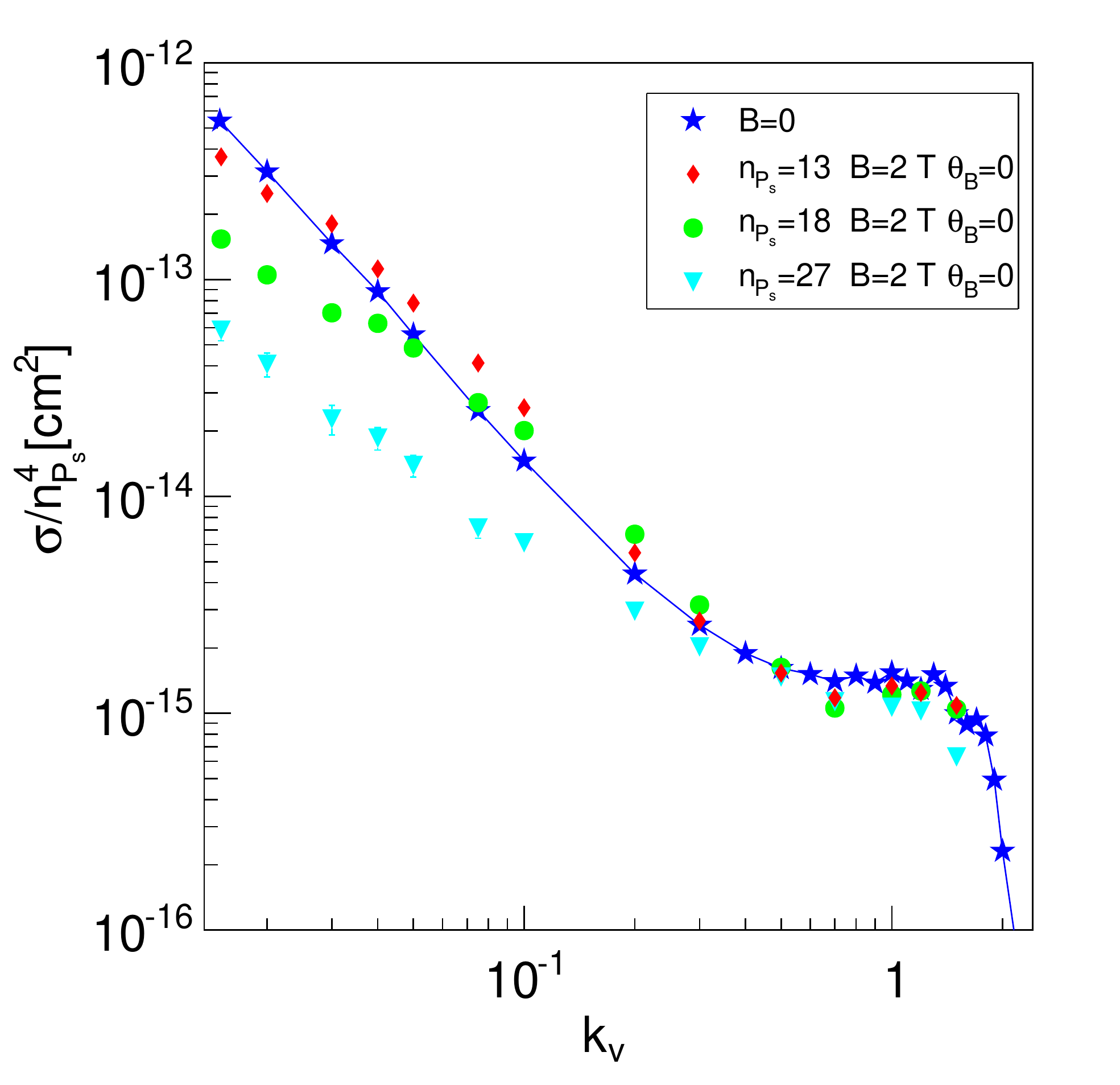}}
\caption{Charge exchange cross section divided for $n_{P_s}^4$ as a function of $k_v$ calculated with B=2 T, $\Theta_B=0$ and with  $n_{P_s}$=13, 18, 27.
 For comparison the normalised zero magnetic field cross section is reported (blue stars). Compare with figure \ref{fig:AllB0}.}
\label{fig:B2TVariousn}
\end{center}
\end{figure} 

\begin{figure}
\begin{center}
\centering{\includegraphics[width = 0.5 \textwidth]{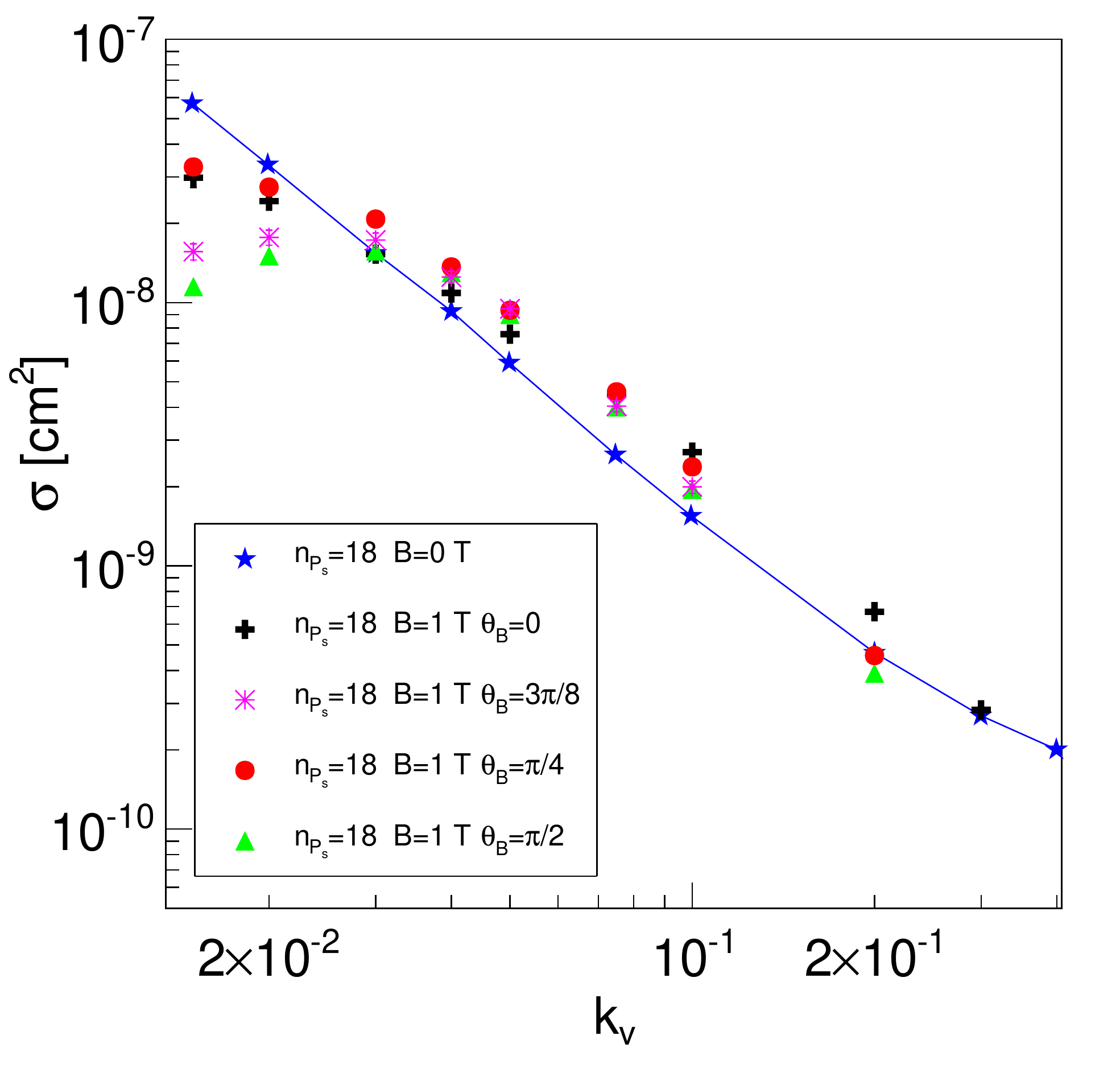}}
\caption{Charge exchange cross section calculated with B=1 T for positronium with principal quantum number $n_{P_s}$=18 and four values of the angle $\theta_B$. For comparison the zero magnetic field cross section is reported (blue stars).}
\label{fig:nPs18B1}
\end{center}
\end{figure}

\begin{figure}
\begin{center}
\centering{\includegraphics[width = 0.5 \textwidth]{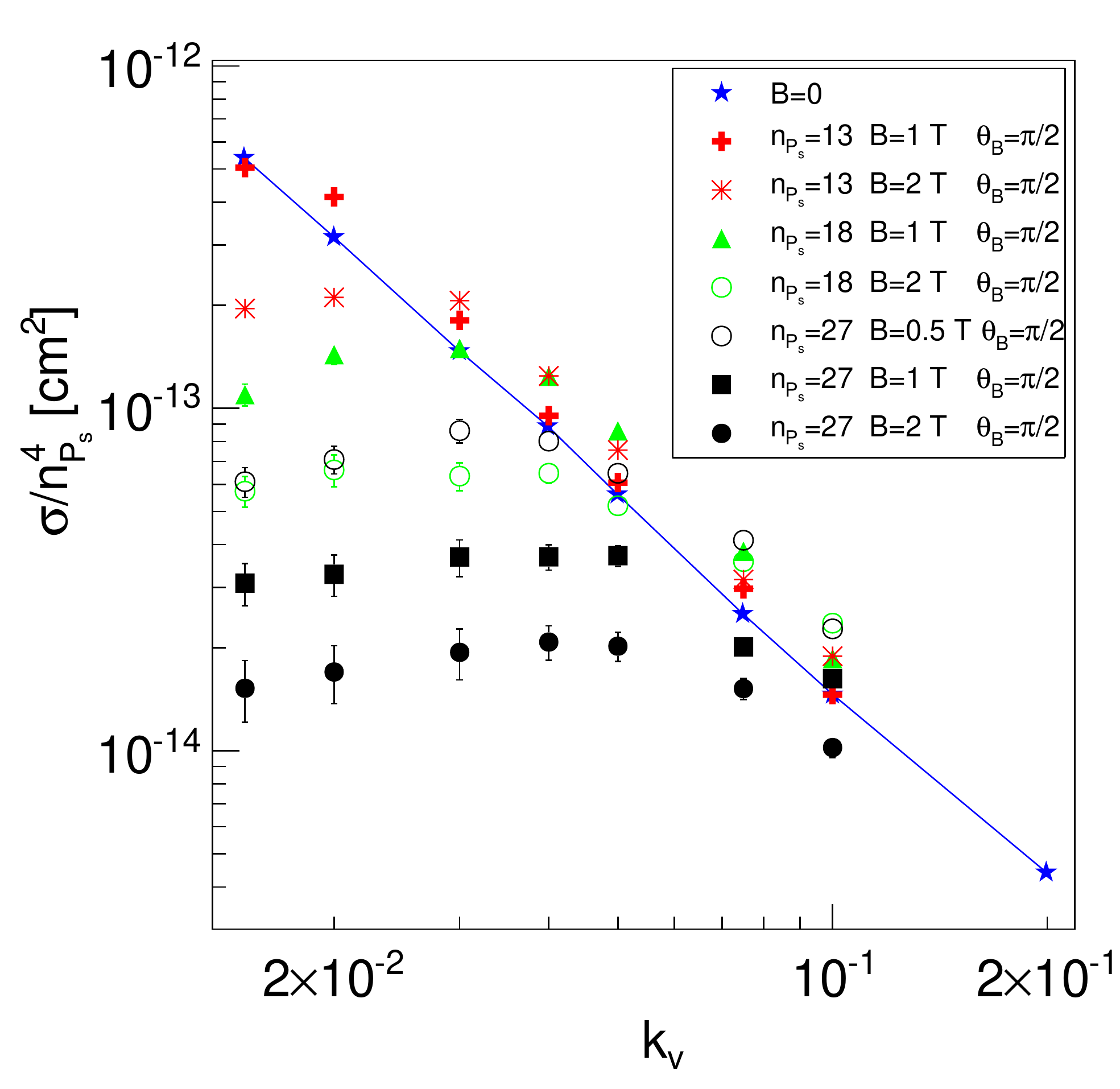}}
\caption{Low velocity charge exchange cross section calculated with interesting combinations of B=0.5, 1, 2 T  and  $n_{P_s}$=13, 18, 27  and  $\theta_B= \pi/2$. For comparison the zero magnetic field cross section is reported (blue stars).}
\label{fig:nPsVariousB}
\end{center}
\end{figure}

\begin{figure*}
\begin{center}
\centering{\includegraphics[width = 1 \textwidth]{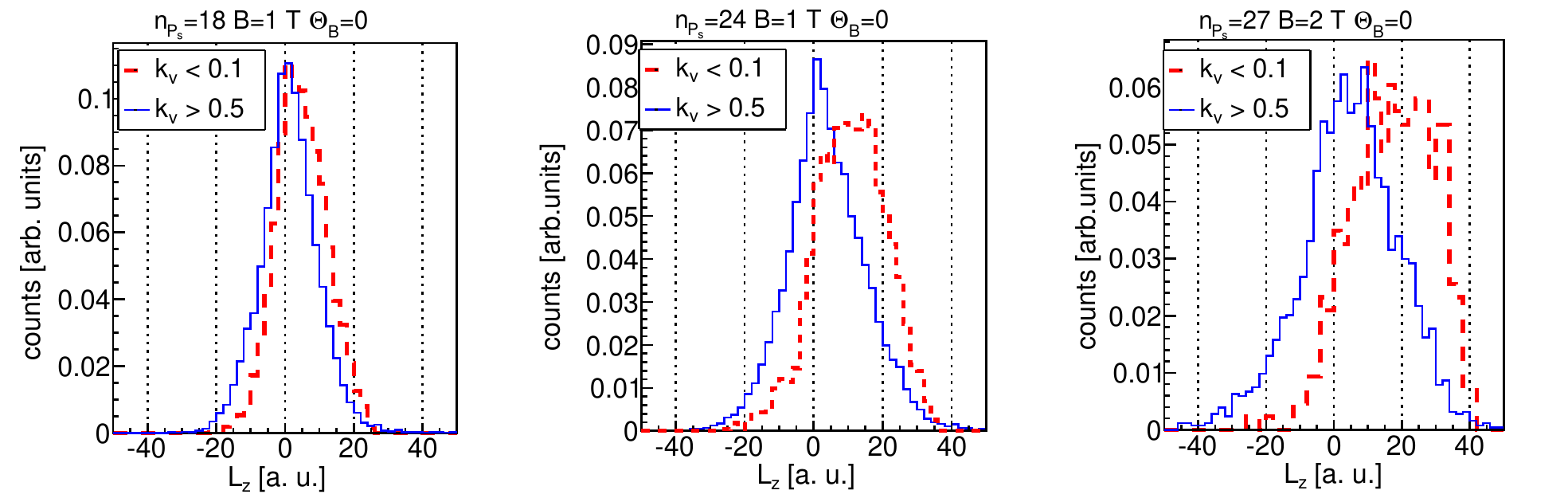}}
\caption{Distributions of $L_z$ of antihydrogen produced with magnetic field along $z$. Without magnetic field the distribution is symmetric.}
\label{fig:Lz}
\end{center}
\end{figure*}  

We have calculated the charge exchange cross section for some values of magnetic field of interest in antihydrogen experiments and for some reference values of $n_{Ps}$.
We  have also studied the effect of the angle $\Theta_B$ of the magnetic field with respect to the positronium flight direction ($z$ axis). 

Figures \ref{fig:B1TVariousn}  and  \ref{fig:B2TVariousn} compare  the zero magnetic field 
 cross section normalised to $n_{P_s}^4$     to the same quantity  obtained for some reference values of $n_{P_s}$ and  two values of $B$ (1 T and   2 T) with $\Theta_B=0$. 
 Note that the magnetic field not only breaks the universality of the shape of the normalised cross section as a function of $k_v$ shown in figure \ref{fig:AllB0} but also it  destroys the $1/k_v^2$ law.
 The curves describing $\sigma/n_{P_s}^4$ versus $k_v$ in magnetic field cross the reference zero field curve when $k_v= k_v^X$. In figure \ref{fig:B1TVariousn}, 
 as example, $k_v^X \simeq 0.1 (0.03)  (0.02) $ if $n_{P_s}=27 (18)  (16) $ while if the $n_{P_s}=13$ then $k_v^X$ is lower than the values reported in the plot.
 There is an interesting  range of positronium velocity satisfying    $k_v>k_v^X$  where   the charge exchange cross section in magnetic field is slightly higher than that in 
 absence of field. However if $k_v<k_v^X$ then $\sigma/n_{P_s}^4$ in presence of magnetic field significantly deviates from the $1/k_v^2$ law and it reaches values lower than the corresponding field free ones.
 The comparison of the results of figures \ref{fig:B1TVariousn} and \ref{fig:B2TVariousn}  indicates that $k_v^X$  is a function of both B and $n_{P_s}$.
 
The low velocity reduction of the cross section also depends on the angle $\Theta_B$ thus making the $k_v^X$ value also a function of $\Theta_B$   with the maximum reduction with respect to the field free case obtained when positronium  flies perpendicular to the field. To our knowledge this is the first time that  directional effects in  charge exchange collisions in magnetic field  are singled out.  The dependance upon the angle  clearly appears in 
 figure \ref{fig:nPs18B1} which compares the field free cross section as a function of $k_v$ for $n_{P_s}=18$ with that obtained with B=1 T and some $\Theta_B$ values.
 Figure \ref{fig:nPsVariousB} compares the low velocity scaled cross section calculated with worst case angle $\theta_B= \pi/2$ and some values of magnetic field and $n_{P_s}$ and it shows how the deviation from  the $1/k_v^2$ regime is influenced by these parameters. 

\subsection{Asymmetry of the distribution of $\bar H$ angular momentum} 
The distribution of the principal quantum numbers of the antihydrogen and its velocity are not affected in relevant way  by the magnetic field. 
As already stated in \cite{TwoStage}, the magnetic field influences the distribution of the component of the angular momentum in the direction of the field and it favours the formation of antihydrogen in high field seeking states. Despite of the slightly lower values of the magnetic field here considered, we obtain a result similar to that of \cite{TwoStage}
but our analysis as a function of the velocity of the incoming positronium shows that the effect is velocity dependent and it is strongly pronounced  for $k_v $
values  corresponding to the $1/k_v^2$ regime of the field free cross section.  Figure \ref{fig:Lz} compares some examples of  distributions of the $z$ component of the canonical angular momentum $L_z= (x_{\bar H} v_{y\bar H} - y_{\bar H} v_{x \bar H} ) + B ( x^2_{\bar H} + y^2_{\bar H})$ of the antihydrogen formed with $\vec B$ directed along $z$ for high and low velocity of the positronium: though the effect depends on $n_{P_s}$ and $B$ in general the asimmetry of the $L_z$ distribution toward positive value is reduced if the collision velocity increases.
The field free distribution is   symmetric.

\section{Conclusions}
The charge exchange reaction  between Rydberg positronium and cold antiprotons is of high interest as it offers the possibility to obtain  cold antiatoms being  no energy  externally supplied to the antiprotons during the formation process. In fact it can be  experimentally implemented by preparing cold antiprotons in a trap and then letting the Rydberg positronium fly through them \cite{AEgIS}. The temperature of the resulting antihydrogen is thus  limited only by the antiproton temperature before the reaction and by the recoil energy.
On the contrary in the antihydrogen formation by three-body recombination,  the electrically charged antiprotons and positrons must be trapped using nested traps \cite{nestedtrap1} and antiprotons
are gently launched through the previously cooled positrons cloud. As result, antihydrogen is typically formed with energies higher than the the positron thermal energy because the 
antiprotons do not thermalize before the capture \cite{TwoStage}.
High Rydberg states of positronium are preferred in the charge exchange process  as the cross section is proportional to   $n_{P_s}^4$ and the recoil energy decreases while   $n_{P_s}$ increases.
The results
  here reported suggest that  charge exchange  with
$n_{Ps}$ in the range 13-20, as foreseen for example  in the AEgIS experiment, is a very effective channel for the  production of antihydrogen with kinetic energy corresponding to about 100 mK or below.

In absence of  magnetic field, when the $P_s$ centre of mass velocity is reduced  below  about  0.2 -0.3 times the classical  velocity of the positron in the circular Kepler orbit, the cross section raises as $1/E_{P_s}^{cm}$. This work shows for the first time this behavior for Rydberg states of positronium. 
It is also interesting  that the CTMC method gives results in fair  agreement with  the CCC quantum model for low $n_{P_s}$ values ($n_{P_s}<3$) in the low velocity collision regime.

The low velocity behavior  of the cross section is of high experimental interest and in fact efforts are already ongoing for producing cold positronium  \cite{PsProduction}.
In the AEgIS  scenario antiprotons are trapped and cooled in a  Penning-Malmberg  trap and  antihydrogen is produced when Rydberg positronium traverse the antiproton cloud.
Positronium atoms are  formed by launching  positrons  towards a nanoporous target material where they lose their energy and bound with high probability to an electron.
Positronium atoms  cool by collisions with the pore walls until eventually they reach thermal equilibrium with the target. Once they emerge in vacuum  they are    excited   to selected Rydberg states by laser pulses \cite{AEgISn3}.
The velocity  of the emerging positronium can be taylored  by a proper selection  of the materials used to build the target  together with the optimisation  of its temperature and  properties and finally by a suitable choice of the positron implantation energy \cite{PsProduction}.
With $n_{P_s}$ in the range 13-20 the onset of the $1/E_{P_s}^{cm}$ regime  is around 1 meV which represents an energy value well reachable with cryogenic (10 K) positronium  formation targets. Colder targets can be operated with consequent slower emitted positronium. 
Progresses toward laser cooling of positronium outside the target 
\cite{PsLaserCooling}  \cite{PsLaserCooling1} are of great interest as a further method to obtain large samples of very cold $P_s$.

A magnetic field is unavoidable in the present experimental  antihydrogen setup as  it is needed to trap and manipulate antiprotons. Our CTMC studies   showed that in presence of magnetic field there is a value of $ k_v^X$ below which the cross section
does not follow  $1/E_{P_s}^{cm}$ law when the $P_s$ velocity is reduced. When  $k_v<k_v^X$ the cross section in presence of magnetic field is lower than the field free one.
$k_v^X$  depends on the value of B, $n_{P_s}$, $\Theta_B$.
Our results show that the effect   is tolerable  if the magnetic field is kept around 1 T or below as expected in \cite{AEgIS} and if $n_{P_s}<18-20$.
Particularly interesting  is also  the increase of the charge exchange cross section for $k_v>k_v^X$.

We showed for the first time that the dynamics of charge exchange process is affected  by the
 the angle $\Theta_B$ between the magnetic field and the flight direction of the incoming positronium. This effect produces a dependance of the cross section upon $\Theta_B$  which is   significant  even at moderate magnetic fields of 1-2 T; our results suggest that these effects should  become more prominent at higher field values.

\par
\par
This work has been funded by INFN (Italy).

\end{document}